\documentclass[prl,twocolumn,superscriptaddress,nofootinbib,noshowpacs,preprintnumbers,longbibliography,floatfix]{revtex4-2}
\usepackage{blindtext}
\usepackage{tikz}
\usepackage{dsfont}
\usepackage{tikz-cd}
\usepackage{amsmath,amssymb}
\usepackage{graphicx}
\usepackage{cleveref}
\usepackage{braket}
\usetikzlibrary{decorations.markings}
 \usepackage{subcaption}
\usepackage[font=small,labelfont=bf,
   justification=Justified,
   format=plain]{caption}
  \usepackage{MnSymbol}

\usepackage{ifthen}
\usepackage{pgfplots}
\pgfplotsset{compat=newest}
\usetikzlibrary{plotmarks}
\usetikzlibrary{arrows.meta}
\usepgfplotslibrary{patchplots}
\usepackage{grffile}
\usetikzlibrary[pgfplots.groupplots]
\pgfplotsset{every axis/.append style={
                    label style={font=\scriptsize},
                    tick label style={font=\scriptsize},
                    legend style={font=\scriptsize},
                    legend image post style={scale=0.5}
                    }
              }
\pgfplotsset{minor grid style={dotted,gray}} 
\pgfplotsset{major grid style={dashed,gray}}
\pgfplotsset{every axis/.append style={thick, tick style=semithick}}
\pgfplotsset{/pgf/number format/1000 sep={}}
\pgfplotsset{xticklabel style={/pgf/number format/fixed},yticklabel style={/pgf/number format/fixed}}

\newlength\figureheight
\newlength\figurewidth
\DeclareUnicodeCharacter{2212}{−}
\usepgfplotslibrary{groupplots,dateplot}
\usetikzlibrary{patterns,shapes.arrows}
\usepgfplotslibrary[groupplots,dateplot]
\usetikzlibrary[patterns,shapes.arrows]

\newcommand{\h}[1]{\hat{#1}}
\newcommand{\muc}{\mu_\mathrm{c}}

\begin{document}
\title{Crystalline phases at finite winding densities in a quantum link ladder}
\author{Paolo Stornati}
\email{paolo.stornati@icfo.eu}
\affiliation{ICFO-Institut de Ciencies Fotoniques, The Barcelona Institute of 
Science and Technology, Av. Carl Friedrich Gauss 3, 08860 Castelldefels (Barcelona), Spain}

\author{Philipp Krah}
\affiliation{TU Berlin,
    Institute of Mathematics,
    Straße des 17. Juni 136, 10623 Berlin, Germany}

\author{Karl Jansen}
\affiliation{Deutsches Elektronen-Synchrotron DESY,Platanenallee 6, 15738 Zeuthen, Germany}
\author{Debasish Banerjee}
\affiliation{Theory Division, Saha Institute of Nuclear Physics, 1/AF Bidhan Nagar, Kolkata 700064, India}
\affiliation{Homi Bhabha National Institute, Training School Complex, Anushaktinagar, Mumbai 400094,India}

\begin{abstract}
Condensed matter physics of gauge theories coupled to fermions can exhibit a rich
phase structure, but are nevertheless very difficult to study in Monte Carlo simulations
when they are afflicted by a sign problem.
As an alternate approach, we use tensor network methods to explore the finite density 
physics of Abelian gauge theories without dynamical matter. As a concrete example, we 
consider the $U(1)$ gauge invariant quantum link ladder with spin-$\frac{1}{2}$ gauge 
fields in an external electric field which cause the winding electric fluxes to condense 
in the ground state. We demonstrate how the electric flux tubes arrange themselves in 
the bulk giving rise to crystalline patterns, whose period can be controlled by tuning 
the external field. We propose observables to detect the transitions in ground state
properties not only in numerical experiments, but also in future cold-atom realizations. A systematic procedure for reaching 
the thermodynamic limit, as well as extending the studies from ladders to
extended geometries is outlined.
\end{abstract}

\maketitle
\paragraph{Introduction.--}
 Finite chemical potentials are expected to 
 give rise to novel phases and correlations otherwise absent 
in the ground state of quantum field theories or quantum many-body systems. Two physically
relevant examples are Quantum Chromodynamics (QCD) and the Hubbard model. Markov Chain Monte Carlo
(MCMC) methods to solve QCD regulated on the lattice can explain 
properties of hadrons, such as their masses, binding energies, and scattering cross-sections. 
At finite baryon densities, $\mu_B$, relevant for e.g., the description of the interior of neutron 
stars or the very early universe, the MCMC methods suffer from the infamous sign problem. The Hubbard 
model, on the other hand, is a pedagogical system
to describe a variety of phases of strongly correlated electrons. At finite doping, it is expected 
to host high-temperature superconducting phases and provide a model for many physically interesting 
materials. Once again, the regime of non-zero doping is difficult 
to investigate numerically using Monte Carlo methods due to the sign problem.

 Finite density physics of scalar and fermionic theories in various space-time dimensions have been
extensively investigated \cite{Banerjee2010, Aarts2010, Katz2016, Bloch2021, Gupta2010, Ayyar2017, Banuls2016}.
We extend such studies which dealt with point particles to pure gauge theories without dynamical 
matter fields containing loop operators. The simplest scenario is an $U(1)$ Abelian lattice
gauge theory in a finite volume and in 2+1 dimensions, where gauge-invariant winding electric flux 
strings can be excited by coupling a chemical potential to each of the global $U(1)$ centre-symmetry 
generators. Each sector is labelled by a set of integers $(\mathbb{Z}_1, \mathbb{Z}_2)$, 
indicating the number of windings in a specified spatial direction. Moreover, these sectors 
are topological in nature, and states in a given winding number sector cannot be  
smoothly deformed to another sector. Further, the electric flux tubes are non-local extended excitations,
unlike the point-like bosonic or fermionic particles, and their properties at finite densities 
could in principle be considerably different. 

 Flux tubes have been played a prominent role in the description of various physical phenomena. 
Nielsen and Olesen \cite{Nielsen1973} introduced the field theory of a vortex-line model,
also identified with dual strings. These are flux tubes, similar to the ones that occur in the theory 
of type-II superconductors, and are responsible for most of the low-energy physics in the strong 
coupling limit. Classical and semi-classical analysis involving electric fluxes interacting with 
a gas of monopoles, giving rise to confinement have been discussed in \cite{Banks1977,Polyakov1976}. 
Non-abelian generalizations of such operators, called disorder operators, were introduced by 
't Hooft to analyse the phases of non-Abelian gauge theories \cite{tHooft1977}. 

  We consider the condensed matter physics of these flux tubes in 2+1-dimensional U(1) gauge theory. 
Previous studies have used the path integral formulation by either exploiting the dual representation 
of Abelian lattice gauge theories \cite{Trottier1993,Zach1997}, or by using the multilevel algorithm 
\cite{Koma2004} and explored properties such as the profile of the electric flux lines connecting 
static charges, or the variation of the potential between two charges with increasing the representation 
of the charges. 
Among other things, this provides valuable insights about the attractive or repulsive nature of the 
flux tubes. 
 
  In this article, we use the Hamiltonian formulation of a $U(1)$ quantum link ladder (QLL)
\cite{Chandrasekharan1996}. This theory is known to have novel crystalline confined phases which 
carry fractional electric flux excitations \cite{Banerjee2013}, possess anomalously localized 
excited states \cite{Banerjee2021}, and are the building blocks of spin-ice compounds 
\cite{Shannon2004,Benton2012}. While it is known how to simulate the theory with an improved 
cluster algorithm at zero and finite temperature \cite{Banerjee2021a}, this method has not
been extended to deal with the scenario at finite winding chemical potential. Instead, we use 
tensor network methods (see for review \cite{Schollwoeck2011}) to perform an ab-initio study of 
the system at finite winding 
density. Thanks to the rapid development of quantum simulators, the key elements for realizing 
this microscopic model on digital and analogue quantum computers are already available
\cite{Celi2020,Paulson2021,Huffman2021}. The finite density physics investigated in the article is
ideal to be observed in a quantum computing setup. The open boundaries and gauge invariance realized 
with quantum spin operators are very natural for quantum simulators. 
\tikzset{middlearrow/.style={
        decoration={markings,
            mark= at position 0.65 with {\arrow[scale=2]{#1}} ,
        },
        postaction={decorate}
    }
}
\begin{figure}[t!]
\centering
\resizebox{0.4\textwidth}{!}{  
\begin{tikzpicture}
\def \dx{2};
\def \dy{2};
\def \nbx{8};
\def \nby{3};
\foreach \x in {3,...,\nbx} {
    \foreach \y in {1,...,\nby} {
        
            \node at (\x*\dx,\y*\dy) [circle, fill=black] {};
        
    }
}
\foreach \x in {3,...,\nbx} {
        \draw (\x*\dx,\dy) -- ( \x*\dx,\nby*\dy);  
}
\foreach \y in {1,...,2} {
    \draw [thick](3*\dx,\y*\dy) -- ( \nbx*\dx,\y*\dy);
}

\foreach \y in {3} {
    \draw [thick, dashed](3*\dx,\y*\dy) -- ( \nbx*\dx,\y*\dy);
}

\draw[->] (2.8*\dx,1.5 ) -- (2.8*\dx+1,1.5) node [pos=0.66,below] {\large$\hat{x}$};
\draw[->](2.8*\dx,1.5 )  -- (2.8*\dx, +\dy+0.5) node [pos=0.66,left] {\large$\hat{y}$};

\node[thick] at (4.5*\dx,1.5*\dy) {\huge$\circlearrowright$};
\def \xO{8};
\def \yO{2};
\draw[thick,middlearrow={latex},draw opacity=0] (\xO+\dx,\yO) -- ( \xO,\yO);
\draw[thick,middlearrow={latex},draw opacity=0] (\xO+\dx,\yO+\dy) -- (\xO+\dx,\yO) ;
\draw[thick,middlearrow={latex},draw opacity=0] ( \xO,\yO+\dy) -- (\xO+\dx,\yO+\dy);
\draw[thick,middlearrow={latex},draw opacity=0] ( \xO,\yO) -- (\xO,\yO+\dy);
\node[thick] at (4.5*\dx,2.5*\dy) {\huge$\circlearrowleft$};
\def \xO{8};
\def \yO{2};
\draw[thick,middlearrow={latex},draw opacity=0] (\xO+\dx,\yO+\dy) -- (\xO+\dx,\yO+2*\dy) ;
\draw[thick,middlearrow={latex},draw opacity=0] ( \xO+\dx,\yO+2*\dy) -- (\xO,\yO+2*\dy);
\draw[thick,middlearrow={latex},draw opacity=0] ( \xO,\yO+2*\dy) -- (\xO,\yO+\dy);

\node[thick] at (6.5*\dx,2.5*\dy) {\huge$\circlearrowright$};
\def \xO{6*\dx};
\def \yO{2*\dy};
\draw[thick,middlearrow={latex},draw opacity=0] ( \xO+\dx,\yO)--(\xO,\yO);
\draw[thick,middlearrow={latex},draw opacity=0] (\xO+\dx,\yO+\dy) -- (\xO+\dx,\yO);
\draw[thick,middlearrow={latex},draw opacity=0] (\xO,\yO+\dy) --(\xO+\dx,\yO+\dy);
\draw[thick,middlearrow={latex},draw opacity=0] (\xO,\yO) --( \xO,\yO+\dy);
\node[thick] at (6.5*\dx,1.5*\dy) {\huge$\ncirclearrowright$};
\def \xO{6*\dx};
\def \yO{1*\dy};
\draw[thick,middlearrow={latex},draw opacity=0] ( \xO,\yO)--(\xO+\dx,\yO);
\draw[thick,middlearrow={latex},draw opacity=0] (\xO+\dx,\yO+\dy) -- (\xO+\dx,\yO);
\draw[thick,middlearrow={latex},draw opacity=0] (\xO,\yO) --( \xO,\yO+\dy);
\draw[line width=1.5pt,dotted] (\xO-0.5*\dx,\yO-0.5\dy) --( \xO-0.5*\dx,\yO+2.5*\dy) node [pos=0.95,right] {\Large$W_y$};
\draw[line width=1.5pt,dotted] (\xO-3.5*\dx,\yO+1.2*\dy) --( \xO+2.8*\dx,\yO+1.2*\dy) node [pos=0.95,above] {\Large$W_x$};
\end{tikzpicture}
}
 \caption{Ladder geometry of the lattice. 
The periodicity in $\hat{y}$ is indicated by the dashed lines. Two flippable plaquettes 
($\circlearrowleft$, $\circlearrowright$) and a non flippable plaquette ($\ncirclearrowright$)
are also shown. The dotted lines indicate the
links which need to be summed to obtain the $x$- and $y$-windings.}
 \label{fig:lattice_conf}
\end{figure}
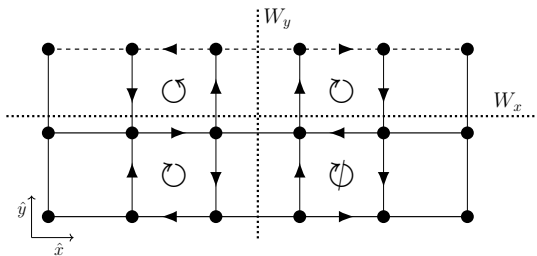

 \paragraph{The $U(1)$ quantum link ladder.--} To illustrate our ideas, we consider the setup of the $U(1)$  
QLL with the gauge fields represented by quantum spins in the spin-$\frac{1}{2}$ representation
on a rectangular lattice $L_x \times L_y$, with $L_y=2$ and $L_x = 6, \dots, 64$, illustrated in 
\Cref{fig:lattice_conf}. Each link degree of freedom has a two-dimensional Hilbert space, and the 
gauge field operator raises (or lowers) the electric flux basis state: 
$U_{r,\h{i}} = S_{r,\h{i}}^+,~U^\dagger_{r,\h{i}} = S_{r,\h{i}}^-,~E_{r,\h{i}} = S_{r,\h{i}}^z$. 
The Hamiltonian consists of two types of plaquette operators: 
\begin{equation}\label{eq:hamiltonian}
    \mathcal H_\square = - J \sum_\square (U_\square + U^\dag_\square ) +  \lambda \sum_\square (U_\square + U^\dag_\square )^2\,,
\end{equation}
where $U_\square= U_{r,\h{i}}U_{r+ \h{i}, \h{j}}U^\dag_{r+\h{j},\h{j}}U^\dag_{r,\h{j}}$. 
One could have added the square of the electric field energy $\sum_{r,\h{i}} E^2_{r,\h{i}}$, 
but for the spin-$\frac{1}{2}$ representation, this is a trivial constant and can be neglected. 
As shown in \Cref{fig:lattice_conf}, the first operator flips any flippable plaquette, while the second 
operator counts the total number of flippable plaquettes. Only two of the 16 states on a 
plaquette are non-trivially acted upon by the plaquette operators. The reduction in the 
number of physical states is due to a local $U(1)$ symmetry, generated by the Gauss law 
\begin{equation}
G_r = \sum_{\h{i}=\h{x},\h{y}} \left( E_{r - \h{i},\h{i}}-E_{r,\h{i}} \right) = 
\sum_{\h{i}=\h{x},\h{y}} \left( S^z_{r-\h{i},\h{i}}-S^z_{r,\h{i}} \right).
\end{equation}
 Physical states satisfy $G_r \ket{\psi} = 0$, which implies the absence of any charge on the lattice.
 In addition, the model has several global symmetries: the lattice translation symmetry 
(by one lattice spacing), the reflection and the rotation symmetry. In addition, there is the
$Z_2$ charge conjugation symmetry: $U \rightarrow U^\dagger, E \rightarrow -E$. However, 
the main object of our interest are the $U(1)^2$ global winding number symmetries, generated
by the operators: 
\begin{equation}
    \label{sec:winding_number}
  W_x = \frac{1}{2 L_y}\sum_{r} S^z_{r,\h{y}} \quad \text{and} \quad  
  W_y = \frac{1}{2 L_x}\sum_{r} S^z_{r,\h{x}}.
\end{equation}
where the sum over $r$ runs over all lattice sites. These operators commute with the Hamiltonian 
and thus classify the eigenstates in terms of
the number of times the flux loops wind the system either along the $x$- or the $y$-direction. 
Therefore, it is natural to couple chemical potentials with strengths $\mu_{x}, \mu_{y}$ to the
Hamiltonian and extend the full Hamiltonian as: $\mathcal H = {\mathcal H}_\square - \mu_{x} W_x - \mu_{y} W_y$.
  
  The windings $W_{x,y}$ are good quantum numbers for periodic boundary conditions.
However, for using open boundary conditions (as we impose in the longer directions, since
we use matrix product states (MPS) in our calculations), one can show that an external
field $(h_x, h_y)$ that couples to the $x$-links and $y$-links respectively serves 
the same purpose, keeping $W_{x,y}$ to be good quantum numbers. With the external field, 
there is a non-trivial contribution from the kinetic energy
term: $\sum_{r,\h{x}} (E_{r,\h{x}} - h_x)^2 + \sum_{r,\h{y}} (E_{r,\h{y}} - h_y)^2 
= - h_x \sum_{r,\h{x}} E_{r,\h{x}} - h_y \sum_{r,\h{y}} E_{r,\h{y}} + {\rm const} 
= - 2 h_x W_y - 2 h_y W_x + {\rm const}$, which is equivalent to coupling the system with 
$\mu_{x,y}$. We will use the latter notation for the rest of the article. 

\begin{figure}[htp!]
    \setlength{\figureheight}{0.7\linewidth}
    \setlength{\figurewidth}{1\linewidth}
    \input{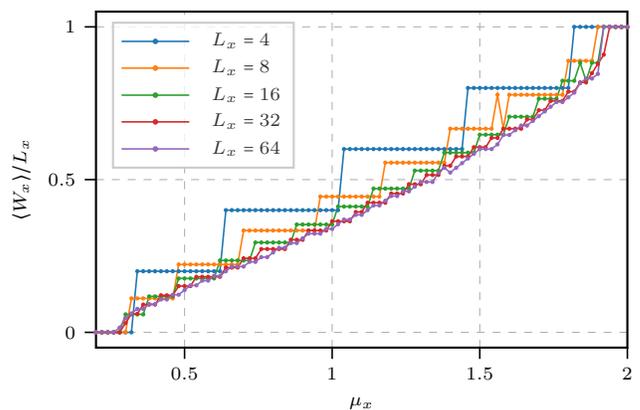}\\
\caption{Staircase structure of the winding numbers $\langle W_x\rangle$ with increasing $\mu$. The 
 plateaux correspond to ground states where the winding flux remains fixed as $\mu$ is varied.
 In the thermodynamic limit, the curve becomes continuous. For large $\mu$, the curve saturates
 just as for fermion (or hard-core bosons). }
\label{fig:delta_winding}
\end{figure}

\paragraph{Numerical methods.--} We begin by noting that the model considered here
a rich ground state phase diagram \cite{Banerjee2013} and realizes novel crystalline confined phases. 
The physics of excited states have revealed the existence of quantum scar states, and atypical real-time 
dynamics \cite{Banerjee2021}. While the former used an efficient cluster Monte-Carlo algorithm,
the latter used large scale exact diagonalization (ED). In this work, we aim to go for system 
sizes beyond the reach of ED, but efficient algorithms at finite $\mu$ are non-trivial to 
construct. While the existing cluster algorithms can update all sectors at finite temperatures,
it is unclear on how to extend this algorithm for finite $\mu$. Therefore, we use density matrix 
renormalization group (DMRG) on MPS states to simulate the ground state phases with increasing 
values of $\mu =\sqrt{\mu_x^2+\mu_y^2}$. The $\mu_y = 0$ is kept throughout the calculations to ensure that 
there is no condensation of strings in the $y$-direction.

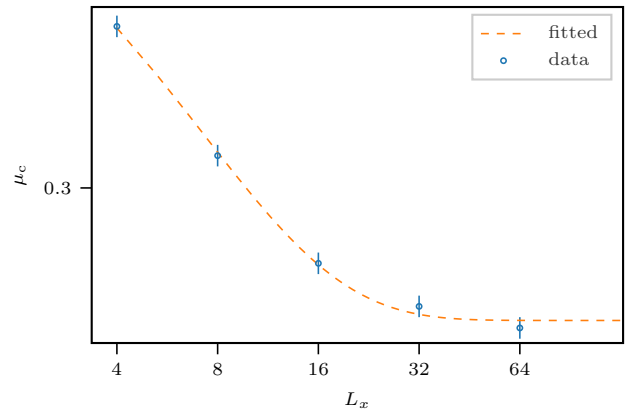
\begin{figure}[htp!]
    \centering
    \setlength{\figureheight}{0.7\linewidth}
    \setlength{\figurewidth}{1\linewidth}
\begin{tikzpicture}

\definecolor{color0}{rgb}{0.12156862745098,0.466666666666667,0.705882352941177}
\definecolor{color1}{rgb}{1,0.498039215686275,0.0549019607843137}

\begin{axis}[
height=\figureheight,
legend cell align={left},
legend style={fill opacity=0.8, draw opacity=1, text opacity=1, draw=white!80!black},
log basis x={10},
minor ytick={},
tick align=outside,
tick pos=left,
width=\figurewidth,
x grid style={white!69.0196078431373!black},
xlabel={\(\displaystyle L_x\)},
xmin=3.36358566101486, xmax=130.218510720348,
xmode=log,
xtick style={color=black},
xtick={4,8,16,32,64},
y grid style={white!69.0196078431373!black},
ylabel={\(\displaystyle \mu_\mathrm{c}\)},
ymin=0.264, ymax=0.342,
ytick style={color=black},
ytick={0.2,0.25,0.3,0.35,0.4},
legend image post style={scale=2},
xticklabels={4,8,16,32,64}
]
\path [draw=color0, semithick]
(axis cs:4,0.335)
--(axis cs:4,0.34);

\path [draw=color0, semithick]
(axis cs:8,0.305)
--(axis cs:8,0.31);

\path [draw=color0, semithick]
(axis cs:16,0.28)
--(axis cs:16,0.285);

\path [draw=color0, semithick]
(axis cs:32,0.27)
--(axis cs:32,0.275);

\path [draw=color0, semithick]
(axis cs:64,0.265)
--(axis cs:64,0.27);

\addplot [semithick, color1, dashed]
table {%
4 0.3371447463244
5.25252525252525 0.326346660677381
6.50505050505051 0.317265514711895
7.75757575757576 0.309628308012277
9.01010101010101 0.303205448342983
10.2626262626263 0.297803849570816
11.5151515151515 0.293261127045683
12.7676767676768 0.289440715939495
14.020202020202 0.28622776578911
15.2727272727273 0.283525687823695
16.5252525252525 0.281253251281128
17.7777777777778 0.279342141421942
19.030303030303 0.277734905829016
20.2828282828283 0.276383227253998
21.5353535353535 0.275246471088192
22.7878787878788 0.274290463791499
24.040404040404 0.273486465556152
25.2929292929293 0.272810306321094
26.5454545454545 0.272241659163608
27.7979797979798 0.271763429224635
29.0505050505051 0.271361239797468
30.3030303030303 0.271023000130432
31.5555555555556 0.270738541950706
32.8080808080808 0.270499313782342
34.0606060606061 0.270298123868965
35.3131313131313 0.270128923972809
36.5656565656566 0.269986627550594
37.8181818181818 0.269866956840179
39.0707070707071 0.269766314261059
40.3232323232323 0.269681674262711
41.5757575757576 0.26961049236949
42.8282828282828 0.269550628687762
44.0808080808081 0.269500283575714
45.3333333333333 0.269457943541925
46.5858585858586 0.269422335746285
47.8383838383838 0.269392389735456
49.0909090909091 0.269367205262549
50.3434343434343 0.269346025223605
51.5959595959596 0.269328212897282
52.8484848484849 0.269313232803531
54.1010101010101 0.269300634605815
55.3535353535354 0.269290039572946
56.6060606060606 0.269281129193538
57.8585858585859 0.269273635600816
59.1111111111111 0.269267333519916
60.3636363636364 0.269262033495596
61.6161616161616 0.269257576196771
62.8686868686869 0.269253827626647
64.1212121212121 0.26925067509447
65.3737373737374 0.269248023827774
66.6262626262626 0.269245794123302
67.8787878787879 0.269243918950939
69.1313131313131 0.269242341938634
70.3838383838384 0.269241015677721
71.6363636363636 0.269239900297705
72.8888888888889 0.269238962267663
74.1414141414141 0.269238173388225
75.3939393939394 0.269237509943838
76.6464646464647 0.269236951989815
77.8989898989899 0.269236482752759
79.1515151515152 0.269236088126309
80.4040404040404 0.269235756247076
81.6565656565657 0.269235477137994
82.9090909090909 0.269235242408396
84.1616161616162 0.269235045001762
85.4141414141414 0.269234878983588
86.6666666666667 0.269234739362981
87.9191919191919 0.26923462194262
89.1717171717172 0.269234523192576
90.4242424242424 0.269234440144192
91.6767676767677 0.26923437030084
92.9292929292929 0.269234311562866
94.1818181818182 0.269234262164469
95.4343434343434 0.269234220620619
96.6868686868687 0.269234185682411
97.9393939393939 0.26923415629952
99.1919191919192 0.269234131588629
100.444444444444 0.269234110806871
101.69696969697 0.269234093329497
102.949494949495 0.269234078631096
104.20202020202 0.269234066269802
105.454545454545 0.269234055874003
106.707070707071 0.269234047131179
107.959595959596 0.2692340397785
109.212121212121 0.269234033594926
110.464646464646 0.269234028394567
111.717171717172 0.269234024021085
112.969696969697 0.269234020343005
114.222222222222 0.269234017249755
115.474747474747 0.269234014648344
116.727272727273 0.269234012460568
117.979797979798 0.269234010620657
119.232323232323 0.2692340090733
120.484848484848 0.269234007771979
121.737373737374 0.269234006677573
122.989898989899 0.269234005757182
124.242424242424 0.269234004983137
125.49494949495 0.269234004332168
126.747474747475 0.269234003784706
128 0.269234003324293
};
\addlegendentry{fitted}
\addplot [semithick, color0, mark=o, mark size=1, mark options={solid}, only marks]
table {%
4 0.3375
8 0.3075
16 0.2825
32 0.2725
64 0.2675
};
\addlegendentry{data}
\end{axis}

\end{tikzpicture}
 \caption{Finite size dependence 
 of $\mu_\mathrm{c}(L_x)=a\,\mathrm{exp}(-b L_x)+\mu_\mathrm{c}^\infty$ 
 on $L_x$. From the fit we determine: $\muc^\infty=0.269$. The error bars are the magnitude of the 
 finite step $\Delta_{\mu_x}=0.0025$ 
 taken to identify the phase transition point.}
\label{fig:mu_critical}
\end{figure}

\paragraph{Condensation of strings.--} The effect of increasing $\mu_x$ on various system
sizes is shown in \Cref{fig:delta_winding}. The more familiar examples of condensation phenomena
are known from bosons and fermions, which are point particles. We notice that with flux strings, 
too, one has the "silver blaze" problem, in which the ground state is unaffected by the chemical 
potential until a threshold value $\muc (L_x)$ is reached, after which the vacuum becomes unstable
to the creation of net flux strings periodically winding around $L_y$. 
 On the smaller lattices, one can 
 clearly observe the step-like structure that results, with 
each step indicating the number of winding strings that have condensed in the vacuum. Plotted in 
terms of the winding density, we notice the smooth approach to the thermodynamic limit in the data
for lattices when reaching $L_x = 64$ (see \Cref{fig:delta_winding}). Note in particular that both the
threshold chemical potential, $\muc(L_x)$, at which condensation phenomena starts and the saturation
chemical potential $\mu_s (L_x)$ have well-defined thermodynamic limits. In \Cref{fig:mu_critical} 
we show the behaviour of $\muc(L_x)$ with increasing volume. It is interesting to note that 
the finite volume dependence is very well described with the same formula that governs the 
dependence of a massive particle in finite volume \cite{Luscher1985}. We note that the
step behaviour of magnetization with an external magnetic field at zero temperature is well known for frustrated
spin systems \cite{Honecker2004}. Recently, a similar behaviour has been reported for the 
ladder Rydberg systems \cite{Sarkar2022}.

  While we have demonstrated the thermodynamic limit for $L_y=2$ ladders, more work is essential
to extend the results to other geometries.
In particular, the 2-d system can be thought of as
a sequence of ladders with increasing $L_y$ at each step. At each fixed $L_y$, we can first take 
the $L_x \to \infty$ limit. Thus, strings that are in general non-local in $L_y$ can condense in
an infinitely wide ladder. For a confining theory, increasing $L_y \to \infty$ is expected to 
yield $\mu_c (L_x \rightarrow \infty, L_y)$ that increases linearly with $L_y$. We postpone the 
demonstration of the thermodynamic limit of larger ladders in a future study, and turn to 
understanding the nature of the phases that are realized in the ground states at finite density.

 \paragraph{Crystalline structures.--} As we demonstrate now, once the winding strings start 
condensing in the ground state, they modulate existing crystalline properties. At $\mu_x = 0$ 
and $\lambda = -1$, the ground state breaks both translation invariance and charge conjugation 
spontaneously \cite{Banerjee2013}. The novel feature at finite $\mu_x$ is the repulsion of condensed strings in the $x$-direction and their subsequent arrangement in periodic intervals. 
This necessarily modulates the pattern of electric fields, $E_{r,\hat{i}}$, from the zero density case. 

\begin{figure}[htp!]
    \begin{minipage}[c]{0.9\columnwidth}
    \setlength{\figureheight}{0.4\linewidth}
    \setlength{\figurewidth}{1\linewidth}
    \input{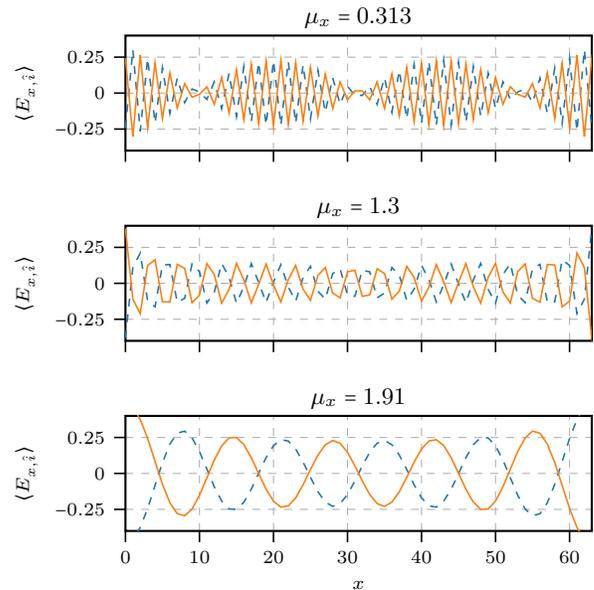}%
    \end{minipage}%
\caption{Vertical electric field, $\braket{E_{x,\hat{i}}}$, for the $L_x = 64$ lattice for three 
different regimes of winding density. The dashed lines correspond to the upper rung and solid lines 
to the electric field on the lower rung of the ladder 
}
\label{fig:electric_field}
\end{figure}

 In \Cref{fig:electric_field}, we show the spatial distribution of the fluxes in the $y$-direction, 
 $E_{x,\hat{i}}$, at three different $\mu_x$ values for the largest lattice $L_x = 64$, representative 
 of three distinct regimes. We call these three different winding regimes: dilute gas regime, half filled 
 and close to saturation regime.
 
 The first regime occurs when the system has just started to condense 
 isolated strings, and the system can be treated as a dilute gas of strings. The top panel of 
 \Cref{fig:electric_field} at $\mu_x = 0.313$ illustrates this case. The three regions where the 
 $\braket{E_{x,\hat{i}}} \approx 0$ marks the location of the winding strings wrapping along the
$y$-direction. We infer that the preference of the strings to stay as far away from each other as 
possible is indicative of their repulsive interaction. Moreover, in between the location of the 
fluxes, the $\braket{E_{x,\hat{i}}}$ displays a regular oscillatory pattern, as also expected for 
$\mu_x = 0$. This arrangement of the fluxes maximizes the total number of flippable plaquettes, as 
preferred by the $\lambda = -1$ term in the Hamiltonian.

 On increasing the filling fraction of the winding density, we notice that the long-wavelength 
 modulations of the electric flux disappear. As shown in the representative middle panel, for 
 $\mu_x = 1.3$, the long-range modulations of $\braket{E_{x,\hat{i}}}$ disappear. The short range 
 oscillations of the horizontal fluxes are still present with twice the period than the previous 
case: the dashed and the solid lines take their maximum positive and negative values ($\approx \pm 0.25$) 
16 times. This regime corresponds to the half-filling of winding strings, now distributed 
evenly through the system, removing traces of previous spatial modulations. Making the system denser 
causes one to approach to the saturation regime, where the electric fields further rearrange to produce 
a smooth coherent oscillation. The bottom panel in \Cref{fig:electric_field}, for $\mu_x = 1.91$ shows 
the coherent oscillations for $L_x = 64$ in this regime, spread over 12-15 lattice spacing. 

 We can also understand the physical properties from the sum of the electric fields on the 
vertical links, $E_{y,\hat{j}}$, which provides the analog of the "particle number density". 
Following our previous discussions, we also expect these profiles to show modulations, 
which are plotted in \Cref{fig:profiles_64a} for our biggest lattice $L_x = 64$. Three distinct 
regimes are also visible in this plot. The set of blue curves represents the regime where the system
has just started to condense isolated strings. It is clear that as $\mu_x$ is slowly increased, the 
winding strings condense in such a way as to maintain maximal separation from each other and the highly 
polarized boundaries. The first such string excitation sits in the middle of the lattice, as shown by 
the maximum in the density profile. The case with three peaks in $E_{x,\hat{j}}$ (at $x = 10a, 30a, 50a$) 
correspond to the profile of $E_{x,\hat{i}}$ at $\mu_x=0.313$ shown in \Cref{fig:electric_field}.
The presence of the fluxes (wrapping vertically) makes the plaquettes non-flippable, which is exactly 
the locations where the horizontal fields are minimum and the vertical fields maximum, demonstrating 
that the strings affect all the local properties. 

On increasing $\mu_x$, the $E_{x,\hat{j}}$ looses the modulations that identify individual fluxes, and a 
smooth distribution, modulated more at the boundaries than in the bulk are visible. Closer to the 
saturation region upon further increasing of $\mu_x$, again longer ranged smooth modulations of the 
"particle density" appear, which now stretch over several lattice spacing. Interestingly, this
length scale seems to be dynamically generated in this regime and rather sensitive to the external
$\mu_x$. The wave number of the oscillations can thus be controlled by tuning the $\mu_x$.

\definecolor{blue0}{rgb}{0.12156862745098,0.466666666666667,0.705882352941177}
\definecolor{orange0}{rgb}{1,0.498039215686275,0.0549019607843137}
\definecolor{green0}{rgb}{0.172549019607843,0.627450980392157,0.172549019607843}

\begin{figure}
    \begin{subfigure}[t]{1\columnwidth}
    \centering
    \setlength{\figurewidth}{0.9\linewidth}
    \setlength{\figureheight}{0.34\linewidth}
    \caption{}
    \input{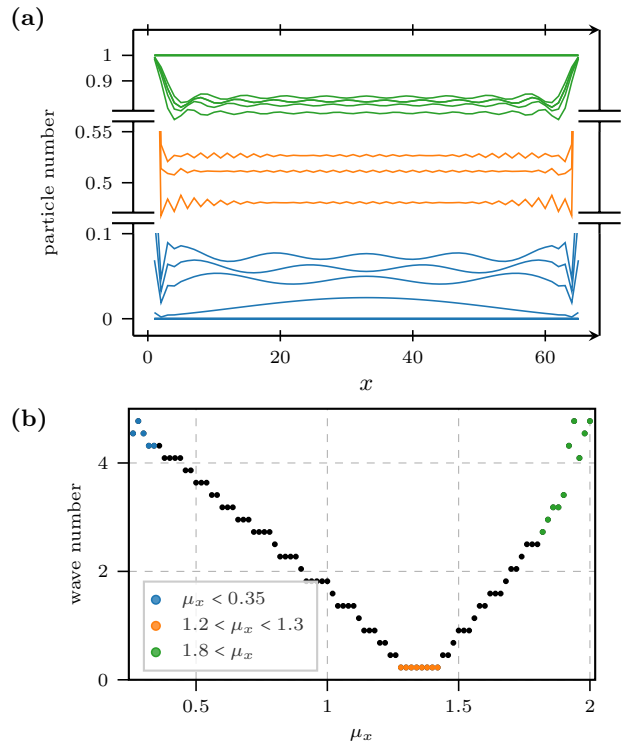}
    \label{fig:profiles_64a}
    \end{subfigure}\\
    \begin{subfigure}[t]{1\columnwidth}
    \centering
    \setlength{\figureheight}{0.6\linewidth}
    \setlength{\figurewidth}{0.9\linewidth}
    \caption{}
    \label{fig:profiles_64b}
\begin{tikzpicture}

\definecolor{color0}{rgb}{0.12156862745098,0.466666666666667,0.705882352941177}
\definecolor{color1}{rgb}{1,0.498039215686275,0.0549019607843137}
\definecolor{color2}{rgb}{0.172549019607843,0.627450980392157,0.172549019607843}

\begin{axis}[
height=\figureheight,
minor xtick={},
minor ytick={},
tick align=outside,
tick pos=left,
width=\figurewidth,
legend style={fill opacity=0.8, draw opacity=1, text opacity=1,
at={(0.03,0.03),mark size=10pt},
legend image post style={scale=4},
anchor=south west,
draw=white!80!black},
x grid style={white!69.0196078431373!black},
xlabel={\(\displaystyle \mu_x\)},
xmajorgrids,
legend cell align={left},
xmin=0.244, xmax=2.02,
xtick style={color=black},
xtick={0,0.5,1,1.5,2},
y grid style={white!69.0196078431373!black},
ylabel={wave number},
ymajorgrids,
ymin=0, ymax=5,
ytick style={color=black},
ytick={0,2,4,6}
]
\addplot [semithick, black, mark=*, mark size=0.8, mark options={solid}, only marks,forget plot]
table {%
0 4.54545454545454
0.02 4.54545454545454
0.04 4.54545454545454
0.06 4.54545454545454
0.08 4.77272727272727
0.1 0.454545454545455
0.12 4.54545454545454
0.14 4.31818181818182
0.16 4.77272727272727
0.18 4.77272727272727
0.2 4.77272727272727
0.22 4.31818181818182
0.24 4.54545454545454
0.26 4.54545454545454
0.28 4.77272727272727
0.3 4.54545454545454
0.32 4.31818181818182
0.34 4.31818181818182
0.36 4.31818181818182
0.38 4.09090909090909
0.4 4.09090909090909
0.42 4.09090909090909
0.44 4.09090909090909
0.46 3.86363636363636
0.48 3.86363636363636
0.5 3.63636363636364
0.52 3.63636363636364
0.54 3.63636363636364
0.56 3.40909090909091
0.58 3.40909090909091
0.6 3.18181818181818
0.62 3.18181818181818
0.64 3.18181818181818
0.66 2.95454545454545
0.68 2.95454545454545
0.7 2.95454545454545
0.72 2.72727272727273
0.74 2.72727272727273
0.76 2.72727272727273
0.78 2.72727272727273
0.8 2.5
0.82 2.27272727272727
0.84 2.27272727272727
0.86 2.27272727272727
0.88 2.27272727272727
0.9 2.04545454545455
0.92 1.81818181818182
0.94 1.81818181818182
0.96 1.81818181818182
0.98 1.81818181818182
1 1.81818181818182
1.02 1.59090909090909
1.04 1.36363636363636
1.06 1.36363636363636
1.08 1.36363636363636
1.1 1.36363636363636
1.12 1.13636363636364
1.14 0.909090909090909
1.16 0.909090909090909
1.18 0.909090909090909
1.2 0.681818181818182
1.22 0.681818181818182
1.24 0.454545454545455
1.26 0.454545454545455
1.28 0.227272727272727
1.3 0.227272727272727
1.32 0.227272727272727
1.34 0.227272727272727
1.36 0.227272727272727
1.38 0.227272727272727
1.4 0.227272727272727
1.42 0.227272727272727
1.44 0.454545454545455
1.46 0.454545454545455
1.48 0.681818181818182
1.5 0.909090909090909
1.52 0.909090909090909
1.54 0.909090909090909
1.56 1.13636363636364
1.58 1.36363636363636
1.6 1.36363636363636
1.62 1.59090909090909
1.64 1.59090909090909
1.66 1.59090909090909
1.68 1.81818181818182
1.7 2.04545454545455
1.72 2.04545454545455
1.74 2.27272727272727
1.76 2.5
1.78 2.5
1.8 2.5
1.82 2.72727272727273
1.84 2.95454545454545
1.86 3.18181818181818
1.88 3.18181818181818
1.9 3.40909090909091
1.92 4.31818181818182
1.94 4.77272727272727
1.96 4.09090909090909
1.98 4.54545454545454
2 4.77272727272727
};
\addplot [semithick, color0, mark=*, mark size=0.8, mark options={solid}, only marks]
table {%
0 4.54545454545454
0.02 4.54545454545454
0.04 4.54545454545454
0.06 4.54545454545454
0.08 4.77272727272727
0.1 0.454545454545455
0.12 4.54545454545454
0.14 4.31818181818182
0.16 4.77272727272727
0.18 4.77272727272727
0.2 4.77272727272727
0.22 4.31818181818182
0.24 4.54545454545454
0.26 4.54545454545454
0.28 4.77272727272727
0.3 4.54545454545454
0.32 4.31818181818182
0.34 4.31818181818182
};
\addlegendentry{$\mu_x<0.35$}
\addplot [semithick, color1, mark=*, mark size=0.8, mark options={solid}, only marks]
table {%
1.28 0.227272727272727
1.3 0.227272727272727
1.32 0.227272727272727
1.34 0.227272727272727
1.36 0.227272727272727
1.38 0.227272727272727
1.4 0.227272727272727
1.42 0.227272727272727
};
\addlegendentry{$1.2<\mu_x<1.3$}
\addplot [semithick, color2, mark=*, mark size=0.8, mark options={solid}, only marks]
table {%
1.82 2.72727272727273
1.84 2.95454545454545
1.86 3.18181818181818
1.88 3.18181818181818
1.9 3.40909090909091
1.92 4.31818181818182
1.94 4.77272727272727
1.96 4.09090909090909
1.98 4.54545454545454
2 4.77272727272727
};
\addlegendentry{$1.8<\mu_x$}
\end{axis}

\end{tikzpicture}
    \end{subfigure}
    \caption{Winding number regimes of the quantum link ladder. (a) The winding number 
    distribution as a function of the distance to one boundary for the three different regimes. 
    (b) The wave number as a function of the chemical potential for the states where the particle 
    number is non-zero or non-saturated. The three different winding regimes are highlighted 
    with colored markers/lines: dilute gas regime ({\color{blue0}$\bullet$}), half filled 
    ({\color{orange0}$\bullet$}) and close to saturation regime ({\color{green0}$\bullet$}). 
    }
    \label{fig:profiles_64}
\end{figure}

\Cref{fig:profiles_64b} shows the wave number of the oscillations as a function of $\mu_x$, 
obtained by identifying the dominant wave-number that contributes in the Fourier transform 
of the vertical electric flux profiles, $E_{x,\h{j}}$. The information in this observable 
is thus the same as in the structure factor up to a global factor, which is given as a 
Fourier transform of  the electric flux correlation function at a particular momentum $k$ (the wave number is $k/2\pi$ in our context). Even in this plot, the aforementioned three regimes in $\mu_x$
are clearly visible. The first non-trivial excitations for small $\mu_x$, present long range 
oscillations whose wave numbers keep decreasing until they saturate to a small value. This is
the regime where the system is approximately half-filled, and for the $L_x = 64$ spans 
from $\mu_x = 1.2, \dots , 1.3$. In this region, the translational invariance is approximately 
recovered. When the chemical potential is increased, the oscillations rise again with a much 
faster rate, as already apparent from the earlier observables. 

\paragraph{Conclusions and Outlook.--} In this Letter, we have explored the phenomenon of 
string condensation in an $U(1)$ Abelian lattice gauge theory realized as a spin-$1/2$ QLM. 
We have demonstrated that our ladder system posses a smooth thermodynamic limit for a fixed $L_y$. 
The system starts to condense strings with the increase in $\mu_x$, and the system exhibits at 
least three different regimes before saturation is reached. Through the profiles of the 
horizontal and vertical electric fluxes, we have shown that the winding strings arrange 
themselves in patterns which behave distinctly in each of the three regimes. In the dilute 
regime, isolated string excitations can be identified, while the half-filled regime is marked 
by an approximate restoration of translation invariance. In the dense region, there is a 
dynamically generated length scale which changes rapidly with $\mu_x$ before the system saturates. Our observables are perfectly suited to be measured in cold atom experiments of 
lattice gauge theory models \cite{cold_athoms_icfo,Martinez:2016yna, Bernien_2017, Schweizer_2019, Mil:2019pbt, Yang:2020yer}.

 There are several directions in which the analysis can be extended. The most obvious is 
to repeat the calculation for larger ladders and study the different regimes that manifest
themselves. Other observables, such as the central charge, and finite size scaling of 
correlation functions could be useful in attempting to understand if there is a phase 
transition between the different regimes. The nature of the origin of the length scale in 
the dense region is also an open question, which might be understood better from an 
effective field theory approach. Another obvious question is if similar phenomena can 
also be observed in QLMs in the spin-$1$ representation, which are very similar to the 
lattice gauge theory formulation by Wilson. 

\paragraph{Acknowledgements.--} We would like to thank Luca Barbiero, Marcello Dalmonte,
Adam Nahum, Arnab Sen, and Uwe-Jens Wiese for helpful discussions. 

PS acknowledges support from: ERC AdG NOQIA; Ministerio de Ciencia y Innovation Agencia Estatal de Investigaciones (PGC2018-097027-B-I00/10.13039/501100011033, CEX2019-000910-S/10.13039/501100011033, Plan National FIDEUA PID2019-106901GB-I00, FPI, QUANTERA MAQS PCI2019-111828-2, QUANTERA DYNAMITE PCI2022-132919, Proyectos de I+D+I “Retos Colaboración” QUSPIN RTC2019-007196-7); European Union NextGenerationEU (PRTR); Fundació Cellex; Fundació Mir-Puig; Generalitat de Catalunya (European Social Fund FEDER and CERCA program (AGAUR Grant No. 2017 SGR 134, QuantumCAT \ U16-011424, co-funded by ERDF Operational Program of Catalonia 2014-2020); Barcelona Supercomputing Center MareNostrum (FI-2022-1-0042); EU Horizon 2020 FET-OPEN OPTOlogic (Grant No 899794); National Science Centre, Poland (Symfonia Grant No.2016/20/W/ST4/00314); European Union’s Horizon 2020 research and innovation programme under the Marie-Skłodowska-Curie grant agreement No 101029393 (STREDCH) and No 847648 (“La Caixa” Junior Leaders fellowships ID100010434: LCF/BQ/PI19/11690013, LCF/BQ/PI20/11760031, LCF/BQ/PR20/11770012, LCF/BQ/PR21/11840013). PK acknowledges support from the Research Training Group ”Differential Equation- and Data-driven 
Models in Life Sciences and Fluid Dynamics: An Interdisciplinary Research Training Group (DAEDALUS)” 
(GRK 2433) funded by the German Research Foundation (DFG).

\bibliography{references}

\begin{thebibliography}{34}%
\makeatletter
\providecommand \@ifxundefined [1]{%
 \@ifx{#1\undefined}
}%
\providecommand \@ifnum [1]{%
 \ifnum #1\expandafter \@firstoftwo
 \else \expandafter \@secondoftwo
 \fi
}%
\providecommand \@ifx [1]{%
 \ifx #1\expandafter \@firstoftwo
 \else \expandafter \@secondoftwo
 \fi
}%
\providecommand \natexlab [1]{#1}%
\providecommand \enquote  [1]{``#1''}%
\providecommand \bibnamefont  [1]{#1}%
\providecommand \bibfnamefont [1]{#1}%
\providecommand \citenamefont [1]{#1}%
\providecommand \href@noop [0]{\@secondoftwo}%
\providecommand \href [0]{\begingroup \@sanitize@url \@href}%
\providecommand \@href[1]{\@@startlink{#1}\@@href}%
\providecommand \@@href[1]{\endgroup#1\@@endlink}%
\providecommand \@sanitize@url [0]{\catcode `\\12\catcode `\$12\catcode
  `\&12\catcode `\#12\catcode `\^12\catcode `\_12\catcode `\%12\relax}%
\providecommand \@@startlink[1]{}%
\providecommand \@@endlink[0]{}%
\providecommand \url  [0]{\begingroup\@sanitize@url \@url }%
\providecommand \@url [1]{\endgroup\@href {#1}{\urlprefix }}%
\providecommand \urlprefix  [0]{URL }%
\providecommand \Eprint [0]{\href }%
\providecommand \doibase [0]{https://doi.org/}%
\providecommand \selectlanguage [0]{\@gobble}%
\providecommand \bibinfo  [0]{\@secondoftwo}%
\providecommand \bibfield  [0]{\@secondoftwo}%
\providecommand \translation [1]{[#1]}%
\providecommand \BibitemOpen [0]{}%
\providecommand \bibitemStop [0]{}%
\providecommand \bibitemNoStop [0]{.\EOS\space}%
\providecommand \EOS [0]{\spacefactor3000\relax}%
\providecommand \BibitemShut  [1]{\csname bibitem#1\endcsname}%
\let\auto@bib@innerbib\@empty
\bibitem [{\citenamefont {Banerjee}\ and\ \citenamefont
  {Chandrasekharan}(2010)}]{Banerjee2010}%
  \BibitemOpen
  \bibfield  {author} {\bibinfo {author} {\bibfnamefont {D.}~\bibnamefont
  {Banerjee}}\ and\ \bibinfo {author} {\bibfnamefont {S.}~\bibnamefont
  {Chandrasekharan}},\ }\bibfield  {title} {\bibinfo {title} {Finite size
  effects in the presence of a chemical potential: A study in the classical
  nonlinear o(2) sigma model},\ }\bibfield  {journal} {\bibinfo  {journal}
  {Physical Review D}\ }\textbf {\bibinfo {volume} {81}},\ \href
  {https://doi.org/10.1103/physrevd.81.125007} {10.1103/physrevd.81.125007}
  (\bibinfo {year} {2010})\BibitemShut {NoStop}%
\bibitem [{\citenamefont {Aarts}\ and\ \citenamefont
  {James}(2010)}]{Aarts2010}%
  \BibitemOpen
  \bibfield  {author} {\bibinfo {author} {\bibfnamefont {G.}~\bibnamefont
  {Aarts}}\ and\ \bibinfo {author} {\bibfnamefont {F.~A.}\ \bibnamefont
  {James}},\ }\bibfield  {title} {\bibinfo {title} {On the convergence of
  complex langevin dynamics: the three-dimensional {XY} model at finite
  chemical potential},\ }\bibfield  {journal} {\bibinfo  {journal} {Journal of
  High Energy Physics}\ }\textbf {\bibinfo {volume} {2010}},\ \href
  {https://doi.org/10.1007/jhep08(2010)020} {10.1007/jhep08(2010)020} (\bibinfo
  {year} {2010})\BibitemShut {NoStop}%
\bibitem [{\citenamefont {Katz}\ \emph {et~al.}(2017)\citenamefont {Katz},
  \citenamefont {Niedermayer}, \citenamefont {Nogradi},\ and\ \citenamefont
  {Torok}}]{Katz2016}%
  \BibitemOpen
  \bibfield  {author} {\bibinfo {author} {\bibfnamefont {S.}~\bibnamefont
  {Katz}}, \bibinfo {author} {\bibfnamefont {F.}~\bibnamefont {Niedermayer}},
  \bibinfo {author} {\bibfnamefont {D.}~\bibnamefont {Nogradi}},\ and\ \bibinfo
  {author} {\bibfnamefont {C.}~\bibnamefont {Torok}},\ }\bibfield  {title}
  {\bibinfo {title} {{Comparison of algorithms for solving the sign problem in
  the O(3) model in 1+1 dimensions at finite chemical potential}},\ }\href
  {https://doi.org/10.1103/PhysRevD.95.054506} {\bibfield  {journal} {\bibinfo
  {journal} {Phys. Rev. D}\ }\textbf {\bibinfo {volume} {95}},\ \bibinfo
  {pages} {054506} (\bibinfo {year} {2017})},\ \Eprint
  {https://arxiv.org/abs/1611.03987} {arXiv:1611.03987 [hep-lat]} \BibitemShut
  {NoStop}%
\bibitem [{\citenamefont {Bloch}\ \emph {et~al.}(2021)\citenamefont {Bloch},
  \citenamefont {Jha}, \citenamefont {Lohmayer},\ and\ \citenamefont
  {Meister}}]{Bloch2021}%
  \BibitemOpen
  \bibfield  {author} {\bibinfo {author} {\bibfnamefont {J.}~\bibnamefont
  {Bloch}}, \bibinfo {author} {\bibfnamefont {R.~G.}\ \bibnamefont {Jha}},
  \bibinfo {author} {\bibfnamefont {R.}~\bibnamefont {Lohmayer}},\ and\
  \bibinfo {author} {\bibfnamefont {M.}~\bibnamefont {Meister}},\ }\bibfield
  {title} {\bibinfo {title} {Tensor renormalization group study of the
  three-dimensional o(2) model},\ }\bibfield  {journal} {\bibinfo  {journal}
  {Physical Review D}\ }\textbf {\bibinfo {volume} {104}},\ \href
  {https://doi.org/10.1103/physrevd.104.094517} {10.1103/physrevd.104.094517}
  (\bibinfo {year} {2021})\BibitemShut {NoStop}%
\bibitem [{\citenamefont {Gupta}(2010)}]{Gupta2010}%
  \BibitemOpen
  \bibfield  {author} {\bibinfo {author} {\bibfnamefont {S.}~\bibnamefont
  {Gupta}},\ }\bibfield  {title} {\bibinfo {title} {{QCD at finite density}},\
  }\href {https://doi.org/10.22323/1.105.0007} {\bibfield  {journal} {\bibinfo
  {journal} {PoS}\ }\textbf {\bibinfo {volume} {LATTICE2010}},\ \bibinfo
  {pages} {007} (\bibinfo {year} {2010})},\ \Eprint
  {https://arxiv.org/abs/1101.0109} {arXiv:1101.0109 [hep-lat]} \BibitemShut
  {NoStop}%
\bibitem [{\citenamefont {Ayyar}\ \emph {et~al.}(2018)\citenamefont {Ayyar},
  \citenamefont {Chandrasekharan},\ and\ \citenamefont
  {Rantaharju}}]{Ayyar2017}%
  \BibitemOpen
  \bibfield  {author} {\bibinfo {author} {\bibfnamefont {V.}~\bibnamefont
  {Ayyar}}, \bibinfo {author} {\bibfnamefont {S.}~\bibnamefont
  {Chandrasekharan}},\ and\ \bibinfo {author} {\bibfnamefont {J.}~\bibnamefont
  {Rantaharju}},\ }\bibfield  {title} {\bibinfo {title} {{Benchmark results in
  the 2D lattice Thirring model with a chemical potential}},\ }\href
  {https://doi.org/10.1103/PhysRevD.97.054501} {\bibfield  {journal} {\bibinfo
  {journal} {Phys. Rev. D}\ }\textbf {\bibinfo {volume} {97}},\ \bibinfo
  {pages} {054501} (\bibinfo {year} {2018})},\ \Eprint
  {https://arxiv.org/abs/1711.07898} {arXiv:1711.07898 [hep-lat]} \BibitemShut
  {NoStop}%
\bibitem [{\citenamefont {Ba\~nuls}\ \emph {et~al.}(2017)\citenamefont
  {Ba\~nuls}, \citenamefont {Cichy}, \citenamefont {Cirac}, \citenamefont
  {Jansen},\ and\ \citenamefont {K\"uhn}}]{Banuls2016}%
  \BibitemOpen
  \bibfield  {author} {\bibinfo {author} {\bibfnamefont {M.~C.}\ \bibnamefont
  {Ba\~nuls}}, \bibinfo {author} {\bibfnamefont {K.}~\bibnamefont {Cichy}},
  \bibinfo {author} {\bibfnamefont {J.~I.}\ \bibnamefont {Cirac}}, \bibinfo
  {author} {\bibfnamefont {K.}~\bibnamefont {Jansen}},\ and\ \bibinfo {author}
  {\bibfnamefont {S.}~\bibnamefont {K\"uhn}},\ }\bibfield  {title} {\bibinfo
  {title} {{Density Induced Phase Transitions in the Schwinger Model: A Study
  with Matrix Product States}},\ }\href
  {https://doi.org/10.1103/PhysRevLett.118.071601} {\bibfield  {journal}
  {\bibinfo  {journal} {Phys. Rev. Lett.}\ }\textbf {\bibinfo {volume} {118}},\
  \bibinfo {pages} {071601} (\bibinfo {year} {2017})},\ \Eprint
  {https://arxiv.org/abs/1611.00705} {arXiv:1611.00705 [hep-lat]} \BibitemShut
  {NoStop}%
\bibitem [{\citenamefont {Nielsen}\ and\ \citenamefont
  {Olesen}(1973)}]{Nielsen1973}%
  \BibitemOpen
  \bibfield  {author} {\bibinfo {author} {\bibfnamefont {H.}~\bibnamefont
  {Nielsen}}\ and\ \bibinfo {author} {\bibfnamefont {P.}~\bibnamefont
  {Olesen}},\ }\bibfield  {title} {\bibinfo {title} {Vortex-line models for
  dual strings},\ }\href
  {https://doi.org/https://doi.org/10.1016/0550-3213(73)90350-7} {\bibfield
  {journal} {\bibinfo  {journal} {Nuclear Physics B}\ }\textbf {\bibinfo
  {volume} {61}},\ \bibinfo {pages} {45} (\bibinfo {year} {1973})}\BibitemShut
  {NoStop}%
\bibitem [{\citenamefont {Banks}\ \emph {et~al.}(1977)\citenamefont {Banks},
  \citenamefont {Myerson},\ and\ \citenamefont {Kogut}}]{Banks1977}%
  \BibitemOpen
  \bibfield  {author} {\bibinfo {author} {\bibfnamefont {T.}~\bibnamefont
  {Banks}}, \bibinfo {author} {\bibfnamefont {R.}~\bibnamefont {Myerson}},\
  and\ \bibinfo {author} {\bibfnamefont {J.}~\bibnamefont {Kogut}},\ }\bibfield
   {title} {\bibinfo {title} {Phase transitions in abelian lattice gauge
  theories},\ }\href
  {https://doi.org/https://doi.org/10.1016/0550-3213(77)90129-8} {\bibfield
  {journal} {\bibinfo  {journal} {Nuclear Physics B}\ }\textbf {\bibinfo
  {volume} {129}},\ \bibinfo {pages} {493} (\bibinfo {year}
  {1977})}\BibitemShut {NoStop}%
\bibitem [{\citenamefont {Polyakov}(1977)}]{Polyakov1976}%
  \BibitemOpen
  \bibfield  {author} {\bibinfo {author} {\bibfnamefont {A.~M.}\ \bibnamefont
  {Polyakov}},\ }\bibfield  {title} {\bibinfo {title} {{Quark Confinement and
  Topology of Gauge Groups}},\ }\href
  {https://doi.org/10.1016/0550-3213(77)90086-4} {\bibfield  {journal}
  {\bibinfo  {journal} {Nucl. Phys. B}\ }\textbf {\bibinfo {volume} {120}},\
  \bibinfo {pages} {429} (\bibinfo {year} {1977})}\BibitemShut {NoStop}%
\bibitem [{\citenamefont {'t~Hooft}(1978)}]{tHooft1977}%
  \BibitemOpen
  \bibfield  {author} {\bibinfo {author} {\bibfnamefont {G.}~\bibnamefont
  {'t~Hooft}},\ }\bibfield  {title} {\bibinfo {title} {{On the Phase Transition
  Towards Permanent Quark Confinement}},\ }\href
  {https://doi.org/10.1016/0550-3213(78)90153-0} {\bibfield  {journal}
  {\bibinfo  {journal} {Nucl. Phys. B}\ }\textbf {\bibinfo {volume} {138}},\
  \bibinfo {pages} {1} (\bibinfo {year} {1978})}\BibitemShut {NoStop}%
\bibitem [{\citenamefont {Trottier}\ and\ \citenamefont
  {Woloshyn}(1993)}]{Trottier1993}%
  \BibitemOpen
  \bibfield  {author} {\bibinfo {author} {\bibfnamefont {H.~D.}\ \bibnamefont
  {Trottier}}\ and\ \bibinfo {author} {\bibfnamefont {R.~M.}\ \bibnamefont
  {Woloshyn}},\ }\bibfield  {title} {\bibinfo {title} {Flux tubes in
  three-dimensional lattice gauge theories},\ }\href
  {https://doi.org/10.1103/physrevd.48.2290} {\bibfield  {journal} {\bibinfo
  {journal} {Physical Review D}\ }\textbf {\bibinfo {volume} {48}},\ \bibinfo
  {pages} {2290} (\bibinfo {year} {1993})}\BibitemShut {NoStop}%
\bibitem [{\citenamefont {Zach}\ \emph {et~al.}(1998)\citenamefont {Zach},
  \citenamefont {Faber},\ and\ \citenamefont {Skala}}]{Zach1997}%
  \BibitemOpen
  \bibfield  {author} {\bibinfo {author} {\bibfnamefont {M.}~\bibnamefont
  {Zach}}, \bibinfo {author} {\bibfnamefont {M.}~\bibnamefont {Faber}},\ and\
  \bibinfo {author} {\bibfnamefont {P.}~\bibnamefont {Skala}},\ }\bibfield
  {title} {\bibinfo {title} {{Flux tubes and their interaction in $U(1)$
  lattice gauge theory}},\ }\href
  {https://doi.org/10.1016/S0550-3213(98)00363-0} {\bibfield  {journal}
  {\bibinfo  {journal} {Nucl. Phys. B}\ }\textbf {\bibinfo {volume} {529}},\
  \bibinfo {pages} {505} (\bibinfo {year} {1998})},\ \Eprint
  {https://arxiv.org/abs/hep-lat/9709017} {arXiv:hep-lat/9709017} \BibitemShut
  {NoStop}%
\bibitem [{\citenamefont {Koma}\ \emph {et~al.}(2004)\citenamefont {Koma},
  \citenamefont {Koma},\ and\ \citenamefont {Majumdar}}]{Koma2004}%
  \BibitemOpen
  \bibfield  {author} {\bibinfo {author} {\bibfnamefont {Y.}~\bibnamefont
  {Koma}}, \bibinfo {author} {\bibfnamefont {M.}~\bibnamefont {Koma}},\ and\
  \bibinfo {author} {\bibfnamefont {P.}~\bibnamefont {Majumdar}},\ }\bibfield
  {title} {\bibinfo {title} {{Static potential, force, and flux-tube profile in
  4D compact $U(1)$ lattice gauge theory with the multi-level algorithm}},\
  }\href {https://doi.org/10.1016/j.nuclphysb.2004.05.024} {\bibfield
  {journal} {\bibinfo  {journal} {Nuclear Physics B}\ }\textbf {\bibinfo
  {volume} {692}},\ \bibinfo {pages} {209} (\bibinfo {year}
  {2004})}\BibitemShut {NoStop}%
\bibitem [{\citenamefont {Chandrasekharan}\ and\ \citenamefont
  {Wiese}(1997)}]{Chandrasekharan1996}%
  \BibitemOpen
  \bibfield  {author} {\bibinfo {author} {\bibfnamefont {S.}~\bibnamefont
  {Chandrasekharan}}\ and\ \bibinfo {author} {\bibfnamefont {U.~J.}\
  \bibnamefont {Wiese}},\ }\bibfield  {title} {\bibinfo {title} {{Quantum link
  models: A Discrete approach to gauge theories}},\ }\href
  {https://doi.org/10.1016/S0550-3213(97)00006-0} {\bibfield  {journal}
  {\bibinfo  {journal} {Nucl. Phys. B}\ }\textbf {\bibinfo {volume} {492}},\
  \bibinfo {pages} {455} (\bibinfo {year} {1997})},\ \Eprint
  {https://arxiv.org/abs/hep-lat/9609042} {arXiv:hep-lat/9609042} \BibitemShut
  {NoStop}%
\bibitem [{\citenamefont {Banerjee}\ \emph {et~al.}(2013)\citenamefont
  {Banerjee}, \citenamefont {Jiang}, \citenamefont {Widmer},\ and\
  \citenamefont {Wiese}}]{Banerjee2013}%
  \BibitemOpen
  \bibfield  {author} {\bibinfo {author} {\bibfnamefont {D.}~\bibnamefont
  {Banerjee}}, \bibinfo {author} {\bibfnamefont {F.-J.}\ \bibnamefont {Jiang}},
  \bibinfo {author} {\bibfnamefont {P.}~\bibnamefont {Widmer}},\ and\ \bibinfo
  {author} {\bibfnamefont {U.-J.}\ \bibnamefont {Wiese}},\ }\bibfield  {title}
  {\bibinfo {title} {The (2 + 1)-d $u(1)$ quantum link model masquerading as
  deconfined criticality},\ }\href
  {https://doi.org/10.1088/1742-5468/2013/12/p12010} {\bibfield  {journal}
  {\bibinfo  {journal} {J. Stat. Mech. Theory Exp.}\ }\textbf {\bibinfo
  {volume} {2013}},\ \bibinfo {pages} {P12010} (\bibinfo {year}
  {2013})}\BibitemShut {NoStop}%
\bibitem [{\citenamefont {Banerjee}\ and\ \citenamefont
  {Sen}(2021)}]{Banerjee2021}%
  \BibitemOpen
  \bibfield  {author} {\bibinfo {author} {\bibfnamefont {D.}~\bibnamefont
  {Banerjee}}\ and\ \bibinfo {author} {\bibfnamefont {A.}~\bibnamefont {Sen}},\
  }\bibfield  {title} {\bibinfo {title} {Quantum scars from zero modes in an
  abelian lattice gauge theory on ladders},\ }\href
  {https://doi.org/10.1103/PhysRevLett.126.220601} {\bibfield  {journal}
  {\bibinfo  {journal} {Phys. Rev. Lett.}\ }\textbf {\bibinfo {volume} {126}},\
  \bibinfo {pages} {220601} (\bibinfo {year} {2021})}\BibitemShut {NoStop}%
\bibitem [{\citenamefont {Shannon}\ \emph {et~al.}(2004)\citenamefont
  {Shannon}, \citenamefont {Misguich},\ and\ \citenamefont
  {Penc}}]{Shannon2004}%
  \BibitemOpen
  \bibfield  {author} {\bibinfo {author} {\bibfnamefont {N.}~\bibnamefont
  {Shannon}}, \bibinfo {author} {\bibfnamefont {G.}~\bibnamefont {Misguich}},\
  and\ \bibinfo {author} {\bibfnamefont {K.}~\bibnamefont {Penc}},\ }\bibfield
  {title} {\bibinfo {title} {Cyclic exchange, isolated states, and spinon
  deconfinement in an $xxz$ heisenberg model on the checkerboard lattice},\
  }\href {https://doi.org/10.1103/PhysRevB.69.220403} {\bibfield  {journal}
  {\bibinfo  {journal} {Phys. Rev. B}\ }\textbf {\bibinfo {volume} {69}},\
  \bibinfo {pages} {220403} (\bibinfo {year} {2004})}\BibitemShut {NoStop}%
\bibitem [{\citenamefont {Benton}\ \emph {et~al.}(2012)\citenamefont {Benton},
  \citenamefont {Sikora},\ and\ \citenamefont {Shannon}}]{Benton2012}%
  \BibitemOpen
  \bibfield  {author} {\bibinfo {author} {\bibfnamefont {O.}~\bibnamefont
  {Benton}}, \bibinfo {author} {\bibfnamefont {O.}~\bibnamefont {Sikora}},\
  and\ \bibinfo {author} {\bibfnamefont {N.}~\bibnamefont {Shannon}},\
  }\bibfield  {title} {\bibinfo {title} {Seeing the light: Experimental
  signatures of emergent electromagnetism in a quantum spin ice},\ }\bibfield
  {journal} {\bibinfo  {journal} {Physical Review B}\ }\textbf {\bibinfo
  {volume} {86}},\ \href {https://doi.org/10.1103/physrevb.86.075154}
  {10.1103/physrevb.86.075154} (\bibinfo {year} {2012})\BibitemShut {NoStop}%
\bibitem [{\citenamefont {Banerjee}(2021)}]{Banerjee2021a}%
  \BibitemOpen
  \bibfield  {author} {\bibinfo {author} {\bibfnamefont {D.}~\bibnamefont
  {Banerjee}},\ }\bibfield  {title} {\bibinfo {title} {Recent progress on
  cluster and meron algorithms for strongly correlated systems},\ }\href
  {https://doi.org/10.1007/s12648-021-02155-5} {\bibfield  {journal} {\bibinfo
  {journal} {Indian Journal of Physics}\ }\textbf {\bibinfo {volume} {95}},\
  \bibinfo {pages} {1669} (\bibinfo {year} {2021})}\BibitemShut {NoStop}%
\bibitem [{\citenamefont {Schollw{\"o}ck}(2011)}]{Schollwoeck2011}%
  \BibitemOpen
  \bibfield  {author} {\bibinfo {author} {\bibfnamefont {U.}~\bibnamefont
  {Schollw{\"o}ck}},\ }\bibfield  {title} {\bibinfo {title} {The density-matrix
  renormalization group in the age of matrix product states},\ }\href@noop {}
  {\bibfield  {journal} {\bibinfo  {journal} {Annals of physics}\ }\textbf
  {\bibinfo {volume} {326}},\ \bibinfo {pages} {96} (\bibinfo {year}
  {2011})}\BibitemShut {NoStop}%
\bibitem [{\citenamefont {Celi}\ \emph {et~al.}(2020)\citenamefont {Celi},
  \citenamefont {Vermersch}, \citenamefont {Viyuela}, \citenamefont {Pichler},
  \citenamefont {Lukin},\ and\ \citenamefont {Zoller}}]{Celi2020}%
  \BibitemOpen
  \bibfield  {author} {\bibinfo {author} {\bibfnamefont {A.}~\bibnamefont
  {Celi}}, \bibinfo {author} {\bibfnamefont {B.}~\bibnamefont {Vermersch}},
  \bibinfo {author} {\bibfnamefont {O.}~\bibnamefont {Viyuela}}, \bibinfo
  {author} {\bibfnamefont {H.}~\bibnamefont {Pichler}}, \bibinfo {author}
  {\bibfnamefont {M.~D.}\ \bibnamefont {Lukin}},\ and\ \bibinfo {author}
  {\bibfnamefont {P.}~\bibnamefont {Zoller}},\ }\bibfield  {title} {\bibinfo
  {title} {Emerging two-dimensional gauge theories in rydberg configurable
  arrays},\ }\href {https://doi.org/10.1103/PhysRevX.10.021057} {\bibfield
  {journal} {\bibinfo  {journal} {Phys. Rev. X}\ }\textbf {\bibinfo {volume}
  {10}},\ \bibinfo {pages} {021057} (\bibinfo {year} {2020})}\BibitemShut
  {NoStop}%
\bibitem [{\citenamefont {Paulson}\ \emph {et~al.}(2021)\citenamefont
  {Paulson}, \citenamefont {Dellantonio}, \citenamefont {Haase}, \citenamefont
  {Celi}, \citenamefont {Kan}, \citenamefont {Jena}, \citenamefont {Kokail},
  \citenamefont {van Bijnen}, \citenamefont {Jansen}, \citenamefont {Zoller},\
  and\ \citenamefont {Muschik}}]{Paulson2021}%
  \BibitemOpen
  \bibfield  {author} {\bibinfo {author} {\bibfnamefont {D.}~\bibnamefont
  {Paulson}}, \bibinfo {author} {\bibfnamefont {L.}~\bibnamefont
  {Dellantonio}}, \bibinfo {author} {\bibfnamefont {J.~F.}\ \bibnamefont
  {Haase}}, \bibinfo {author} {\bibfnamefont {A.}~\bibnamefont {Celi}},
  \bibinfo {author} {\bibfnamefont {A.}~\bibnamefont {Kan}}, \bibinfo {author}
  {\bibfnamefont {A.}~\bibnamefont {Jena}}, \bibinfo {author} {\bibfnamefont
  {C.}~\bibnamefont {Kokail}}, \bibinfo {author} {\bibfnamefont
  {R.}~\bibnamefont {van Bijnen}}, \bibinfo {author} {\bibfnamefont
  {K.}~\bibnamefont {Jansen}}, \bibinfo {author} {\bibfnamefont
  {P.}~\bibnamefont {Zoller}},\ and\ \bibinfo {author} {\bibfnamefont {C.~A.}\
  \bibnamefont {Muschik}},\ }\bibfield  {title} {\bibinfo {title} {Simulating
  2d effects in lattice gauge theories on a quantum computer},\ }\href
  {https://doi.org/10.1103/PRXQuantum.2.030334} {\bibfield  {journal} {\bibinfo
   {journal} {PRX Quantum}\ }\textbf {\bibinfo {volume} {2}},\ \bibinfo {pages}
  {030334} (\bibinfo {year} {2021})}\BibitemShut {NoStop}%
\bibitem [{\citenamefont {Huffman}\ \emph {et~al.}(2021)\citenamefont
  {Huffman}, \citenamefont {Vera},\ and\ \citenamefont
  {Banerjee}}]{Huffman2021}%
  \BibitemOpen
  \bibfield  {author} {\bibinfo {author} {\bibfnamefont {E.}~\bibnamefont
  {Huffman}}, \bibinfo {author} {\bibfnamefont {M.~G.}\ \bibnamefont {Vera}},\
  and\ \bibinfo {author} {\bibfnamefont {D.}~\bibnamefont {Banerjee}},\ }\href
  {https://doi.org/10.48550/ARXIV.2109.15065} {\bibinfo {title} {Real-time
  dynamics of plaquette models using nisq hardware}} (\bibinfo {year}
  {2021})\BibitemShut {NoStop}%
\bibitem [{\citenamefont {Luscher}(1986)}]{Luscher1985}%
  \BibitemOpen
  \bibfield  {author} {\bibinfo {author} {\bibfnamefont {M.}~\bibnamefont
  {Luscher}},\ }\bibfield  {title} {\bibinfo {title} {{Volume Dependence of the
  Energy Spectrum in Massive Quantum Field Theories. 1. Stable Particle
  States}},\ }\href {https://doi.org/10.1007/BF01211589} {\bibfield  {journal}
  {\bibinfo  {journal} {Commun. Math. Phys.}\ }\textbf {\bibinfo {volume}
  {104}},\ \bibinfo {pages} {177} (\bibinfo {year} {1986})}\BibitemShut
  {NoStop}%
\bibitem [{\citenamefont {Honecker}\ \emph {et~al.}(2004)\citenamefont
  {Honecker}, \citenamefont {Schulenburg},\ and\ \citenamefont
  {Richter}}]{Honecker2004}%
  \BibitemOpen
  \bibfield  {author} {\bibinfo {author} {\bibfnamefont {A.}~\bibnamefont
  {Honecker}}, \bibinfo {author} {\bibfnamefont {J.}~\bibnamefont
  {Schulenburg}},\ and\ \bibinfo {author} {\bibfnamefont {J.}~\bibnamefont
  {Richter}},\ }\bibfield  {title} {\bibinfo {title} {Magnetization plateaus in
  frustrated antiferromagnetic quantum spin models},\ }\href
  {https://doi.org/10.1088/0953-8984/16/11/025} {\bibfield  {journal} {\bibinfo
   {journal} {Journal of Physics: Condensed Matter}\ }\textbf {\bibinfo
  {volume} {16}},\ \bibinfo {pages} {S749} (\bibinfo {year}
  {2004})}\BibitemShut {NoStop}%
\bibitem [{\citenamefont {Sarkar}\ \emph {et~al.}(2022)\citenamefont {Sarkar},
  \citenamefont {Pal}, \citenamefont {Sen},\ and\ \citenamefont
  {Sengupta}}]{Sarkar2022}%
  \BibitemOpen
  \bibfield  {author} {\bibinfo {author} {\bibfnamefont {M.}~\bibnamefont
  {Sarkar}}, \bibinfo {author} {\bibfnamefont {M.}~\bibnamefont {Pal}},
  \bibinfo {author} {\bibfnamefont {A.}~\bibnamefont {Sen}},\ and\ \bibinfo
  {author} {\bibfnamefont {K.}~\bibnamefont {Sengupta}},\ }\bibfield  {title}
  {\bibinfo {title} {Quantum order-by-disorder induced phase transition in
  rydberg ladders with staggered detuning},\ }\href@noop {} {\bibfield
  {journal} {\bibinfo  {journal} {arXiv preprint arXiv:2204.12515}\ } (\bibinfo
  {year} {2022})}\BibitemShut {NoStop}%
\bibitem [{\citenamefont {Aidelsburger}\ \emph {et~al.}(2022)\citenamefont
  {Aidelsburger}, \citenamefont {Barbiero}, \citenamefont {Bermudez},
  \citenamefont {Chanda}, \citenamefont {Dauphin}, \citenamefont
  {González-Cuadra}, \citenamefont {Grzybowski}, \citenamefont {Hands},
  \citenamefont {Jendrzejewski}, \citenamefont {Jünemann}, \citenamefont
  {Juzeliūnas}, \citenamefont {Kasper}, \citenamefont {Piga}, \citenamefont
  {Ran}, \citenamefont {Rizzi}, \citenamefont {Sierra}, \citenamefont
  {Tagliacozzo}, \citenamefont {Tirrito}, \citenamefont {Zache}, \citenamefont
  {Zakrzewski}, \citenamefont {Zohar},\ and\ \citenamefont
  {Lewenstein}}]{cold_athoms_icfo}%
  \BibitemOpen
  \bibfield  {author} {\bibinfo {author} {\bibfnamefont {M.}~\bibnamefont
  {Aidelsburger}}, \bibinfo {author} {\bibfnamefont {L.}~\bibnamefont
  {Barbiero}}, \bibinfo {author} {\bibfnamefont {A.}~\bibnamefont {Bermudez}},
  \bibinfo {author} {\bibfnamefont {T.}~\bibnamefont {Chanda}}, \bibinfo
  {author} {\bibfnamefont {A.}~\bibnamefont {Dauphin}}, \bibinfo {author}
  {\bibfnamefont {D.}~\bibnamefont {González-Cuadra}}, \bibinfo {author}
  {\bibfnamefont {P.~R.}\ \bibnamefont {Grzybowski}}, \bibinfo {author}
  {\bibfnamefont {S.}~\bibnamefont {Hands}}, \bibinfo {author} {\bibfnamefont
  {F.}~\bibnamefont {Jendrzejewski}}, \bibinfo {author} {\bibfnamefont
  {J.}~\bibnamefont {Jünemann}}, \bibinfo {author} {\bibfnamefont
  {G.}~\bibnamefont {Juzeliūnas}}, \bibinfo {author} {\bibfnamefont
  {V.}~\bibnamefont {Kasper}}, \bibinfo {author} {\bibfnamefont
  {A.}~\bibnamefont {Piga}}, \bibinfo {author} {\bibfnamefont {S.-J.}\
  \bibnamefont {Ran}}, \bibinfo {author} {\bibfnamefont {M.}~\bibnamefont
  {Rizzi}}, \bibinfo {author} {\bibfnamefont {G.}~\bibnamefont {Sierra}},
  \bibinfo {author} {\bibfnamefont {L.}~\bibnamefont {Tagliacozzo}}, \bibinfo
  {author} {\bibfnamefont {E.}~\bibnamefont {Tirrito}}, \bibinfo {author}
  {\bibfnamefont {T.~V.}\ \bibnamefont {Zache}}, \bibinfo {author}
  {\bibfnamefont {J.}~\bibnamefont {Zakrzewski}}, \bibinfo {author}
  {\bibfnamefont {E.}~\bibnamefont {Zohar}},\ and\ \bibinfo {author}
  {\bibfnamefont {M.}~\bibnamefont {Lewenstein}},\ }\bibfield  {title}
  {\bibinfo {title} {Cold atoms meet lattice gauge theory},\ }\href
  {https://doi.org/10.1098/rsta.2021.0064} {\bibfield  {journal} {\bibinfo
  {journal} {Philos. Trans. R. Soc. A}\ }\textbf {\bibinfo {volume} {380}},\
  \bibinfo {pages} {20210064} (\bibinfo {year} {2022})}\BibitemShut {NoStop}%
\bibitem [{\citenamefont {Martinez}\ \emph {et~al.}(2016)\citenamefont
  {Martinez} \emph {et~al.}}]{Martinez:2016yna}%
  \BibitemOpen
  \bibfield  {author} {\bibinfo {author} {\bibfnamefont {E.~A.}\ \bibnamefont
  {Martinez}} \emph {et~al.},\ }\bibfield  {title} {\bibinfo {title}
  {{Real-time dynamics of lattice gauge theories with a few-qubit quantum
  computer}},\ }\href {https://doi.org/10.1038/nature18318} {\bibfield
  {journal} {\bibinfo  {journal} {Nature}\ }\textbf {\bibinfo {volume} {534}},\
  \bibinfo {pages} {516} (\bibinfo {year} {2016})},\ \Eprint
  {https://arxiv.org/abs/1605.04570} {arXiv:1605.04570 [quant-ph]} \BibitemShut
  {NoStop}%
\bibitem [{\citenamefont {Bernien}\ \emph {et~al.}(2017)\citenamefont
  {Bernien}, \citenamefont {Schwartz}, \citenamefont {Keesling}, \citenamefont
  {Levine}, \citenamefont {Omran}, \citenamefont {Pichler}, \citenamefont
  {Choi}, \citenamefont {Zibrov}, \citenamefont {Endres}, \citenamefont
  {Greiner},\ and\ \citenamefont {et~al.}}]{Bernien_2017}%
  \BibitemOpen
  \bibfield  {author} {\bibinfo {author} {\bibfnamefont {H.}~\bibnamefont
  {Bernien}}, \bibinfo {author} {\bibfnamefont {S.}~\bibnamefont {Schwartz}},
  \bibinfo {author} {\bibfnamefont {A.}~\bibnamefont {Keesling}}, \bibinfo
  {author} {\bibfnamefont {H.}~\bibnamefont {Levine}}, \bibinfo {author}
  {\bibfnamefont {A.}~\bibnamefont {Omran}}, \bibinfo {author} {\bibfnamefont
  {H.}~\bibnamefont {Pichler}}, \bibinfo {author} {\bibfnamefont
  {S.}~\bibnamefont {Choi}}, \bibinfo {author} {\bibfnamefont {A.~S.}\
  \bibnamefont {Zibrov}}, \bibinfo {author} {\bibfnamefont {M.}~\bibnamefont
  {Endres}}, \bibinfo {author} {\bibfnamefont {M.}~\bibnamefont {Greiner}},\
  and\ \bibinfo {author} {\bibnamefont {et~al.}},\ }\bibfield  {title}
  {\bibinfo {title} {Probing many-body dynamics on a 51-atom quantum
  simulator},\ }\href {https://doi.org/10.1038/nature24622} {\bibfield
  {journal} {\bibinfo  {journal} {Nature}\ }\textbf {\bibinfo {volume} {551}},\
  \bibinfo {pages} {579–584} (\bibinfo {year} {2017})}\BibitemShut {NoStop}%
\bibitem [{\citenamefont {Schweizer}\ \emph {et~al.}(2019)\citenamefont
  {Schweizer}, \citenamefont {Grusdt}, \citenamefont {Berngruber},
  \citenamefont {Barbiero}, \citenamefont {Demler}, \citenamefont {Goldman},
  \citenamefont {Bloch},\ and\ \citenamefont {Aidelsburger}}]{Schweizer_2019}%
  \BibitemOpen
  \bibfield  {author} {\bibinfo {author} {\bibfnamefont {C.}~\bibnamefont
  {Schweizer}}, \bibinfo {author} {\bibfnamefont {F.}~\bibnamefont {Grusdt}},
  \bibinfo {author} {\bibfnamefont {M.}~\bibnamefont {Berngruber}}, \bibinfo
  {author} {\bibfnamefont {L.}~\bibnamefont {Barbiero}}, \bibinfo {author}
  {\bibfnamefont {E.}~\bibnamefont {Demler}}, \bibinfo {author} {\bibfnamefont
  {N.}~\bibnamefont {Goldman}}, \bibinfo {author} {\bibfnamefont
  {I.}~\bibnamefont {Bloch}},\ and\ \bibinfo {author} {\bibfnamefont
  {M.}~\bibnamefont {Aidelsburger}},\ }\bibfield  {title} {\bibinfo {title}
  {Floquet approach to $\mathbb{Z}_2$ lattice gauge theories with ultracold
  atoms in optical lattices},\ }\href
  {https://doi.org/10.1038/s41567-019-0649-7} {\bibfield  {journal} {\bibinfo
  {journal} {Nature Physics}\ }\textbf {\bibinfo {volume} {15}},\ \bibinfo
  {pages} {1168–1173} (\bibinfo {year} {2019})}\BibitemShut {NoStop}%
\bibitem [{\citenamefont {Mil}\ \emph {et~al.}(2020)\citenamefont {Mil},
  \citenamefont {Zache}, \citenamefont {Hegde}, \citenamefont {Xia},
  \citenamefont {Bhatt}, \citenamefont {Oberthaler}, \citenamefont {Hauke},
  \citenamefont {Berges},\ and\ \citenamefont {Jendrzejewski}}]{Mil:2019pbt}%
  \BibitemOpen
  \bibfield  {author} {\bibinfo {author} {\bibfnamefont {A.}~\bibnamefont
  {Mil}}, \bibinfo {author} {\bibfnamefont {T.~V.}\ \bibnamefont {Zache}},
  \bibinfo {author} {\bibfnamefont {A.}~\bibnamefont {Hegde}}, \bibinfo
  {author} {\bibfnamefont {A.}~\bibnamefont {Xia}}, \bibinfo {author}
  {\bibfnamefont {R.~P.}\ \bibnamefont {Bhatt}}, \bibinfo {author}
  {\bibfnamefont {M.~K.}\ \bibnamefont {Oberthaler}}, \bibinfo {author}
  {\bibfnamefont {P.}~\bibnamefont {Hauke}}, \bibinfo {author} {\bibfnamefont
  {J.}~\bibnamefont {Berges}},\ and\ \bibinfo {author} {\bibfnamefont
  {F.}~\bibnamefont {Jendrzejewski}},\ }\bibfield  {title} {\bibinfo {title}
  {{A scalable realization of local $U(1)$ gauge invariance in cold atomic
  mixtures}},\ }\href {https://doi.org/10.1126/science.aaz5312} {\bibfield
  {journal} {\bibinfo  {journal} {Science}\ }\textbf {\bibinfo {volume}
  {367}},\ \bibinfo {pages} {1128} (\bibinfo {year} {2020})},\ \Eprint
  {https://arxiv.org/abs/1909.07641} {arXiv:1909.07641 [cond-mat.quant-gas]}
  \BibitemShut {NoStop}%
\bibitem [{\citenamefont {Yang}\ \emph {et~al.}(2020)\citenamefont {Yang},
  \citenamefont {Sun}, \citenamefont {Ott}, \citenamefont {Wang}, \citenamefont
  {Zache}, \citenamefont {Halimeh}, \citenamefont {Yuan}, \citenamefont
  {Hauke},\ and\ \citenamefont {Pan}}]{Yang:2020yer}%
  \BibitemOpen
  \bibfield  {author} {\bibinfo {author} {\bibfnamefont {B.}~\bibnamefont
  {Yang}}, \bibinfo {author} {\bibfnamefont {H.}~\bibnamefont {Sun}}, \bibinfo
  {author} {\bibfnamefont {R.}~\bibnamefont {Ott}}, \bibinfo {author}
  {\bibfnamefont {H.-Y.}\ \bibnamefont {Wang}}, \bibinfo {author}
  {\bibfnamefont {T.~V.}\ \bibnamefont {Zache}}, \bibinfo {author}
  {\bibfnamefont {J.~C.}\ \bibnamefont {Halimeh}}, \bibinfo {author}
  {\bibfnamefont {Z.-S.}\ \bibnamefont {Yuan}}, \bibinfo {author}
  {\bibfnamefont {P.}~\bibnamefont {Hauke}},\ and\ \bibinfo {author}
  {\bibfnamefont {J.-W.}\ \bibnamefont {Pan}},\ }\bibfield  {title} {\bibinfo
  {title} {{Observation of gauge invariance in a 71-site
  Bose\textendash{}Hubbard quantum simulator}},\ }\href
  {https://doi.org/10.1038/s41586-020-2910-8} {\bibfield  {journal} {\bibinfo
  {journal} {Nature}\ }\textbf {\bibinfo {volume} {587}},\ \bibinfo {pages}
  {392} (\bibinfo {year} {2020})},\ \Eprint {https://arxiv.org/abs/2003.08945}
  {arXiv:2003.08945 [cond-mat.quant-gas]} \BibitemShut {NoStop}%
\bibitem [{\citenamefont {Fishman}\ \emph {et~al.}(2020)\citenamefont
  {Fishman}, \citenamefont {White},\ and\ \citenamefont
  {Stoudenmire}}]{itensor}%
  \BibitemOpen
  \bibfield  {author} {\bibinfo {author} {\bibfnamefont {M.}~\bibnamefont
  {Fishman}}, \bibinfo {author} {\bibfnamefont {S.~R.}\ \bibnamefont {White}},\
  and\ \bibinfo {author} {\bibfnamefont {E.~M.}\ \bibnamefont {Stoudenmire}},\
  }\href@noop {} {\bibinfo {title} {The \mbox{ITensor} software library for
  tensor network calculations}} (\bibinfo {year} {2020}),\ \Eprint
  {https://arxiv.org/abs/2007.14822} {arXiv:2007.14822} \BibitemShut {NoStop}%
\end{thebibliography}%
\newpage

\appendix
\section{Supplementary Material}

\paragraph{Numerical Implementation.--}
 For an efficient representation of the ground states in the quantum link ladder,
 we use matrix product states implemented in the \texttt{itensor} software package 
 \cite{itensor}. All the result presented in this work have been extrapolated to 
 infinite bond dimension. During the numerical simulations, the bond dimension $D$ 
 of the matrix product state has been increased up to 1000 ($D_\mathrm{max}$). We 
 have also checked that the final state can be compressed with $D < D_\mathrm{max}$. 
 The stopping criteria used in the optimization procedure is that the difference of 
 the energy after 5 sweeps should be smaller than $10^{-10}$. For the larger volumes 
 and close to the saturation, in the regime of the $\mu_x \sim 1.8$, the numerical 
 simulation becomes unstable. If we run the optimization procedure different times, 
 we might find different the ground state in different topological sectors. This 
 phenomenon is caused because the energy gap between the different topological sectors 
 approaches zero in the thermodynamic limit and close to the saturation regime. Since 
 the gap in this region is smaller than the numerical precision, we could have found 
 inconsistent result for certain parameter values. Nevertheless, we are still able to 
 capture the important properties and extract physical information. 

\paragraph{Observables at $\mu_x=\mu_y = 0$.--}
 In this section, we briefly sketch the physics of the model at zero chemical potential,
but for varying $\lambda$ for completeness. This has already been discussed at length
in \cite{Banerjee2013} for periodic boundary conditions and using an efficient quantum
Monte Carlo algorithm. It was demonstrated in that study that for large negative values
of $\lambda$, where the $J$ term is insignificant, the ground state physics is dominated
by the states with the largest number of flippable plaquettes. This state spontaneously
breaks both the charge conjugation and the lattice translation symmetry. For decreasing
$\lambda$, however, the symmetry breaking pattern changes as the $J$ term increases in
strength. We encounter a phase where the charge conjugation symmetry is restored by
the lattice translation symmetry remains broken. The two phases are connected by a
weak first order phase transition. 

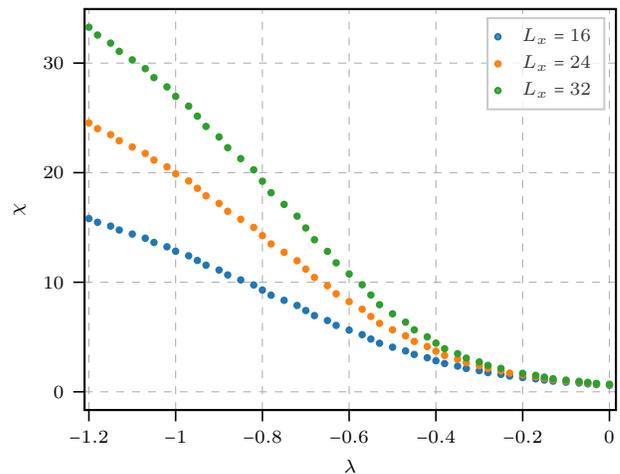
\begin{figure}[htp!]
    \centering
    \setlength{\figureheight}{0.8\linewidth}
    \setlength{\figurewidth}{1.0\linewidth}
\begin{tikzpicture}

\definecolor{color0}{rgb}{0.12156862745098,0.466666666666667,0.705882352941177}
\definecolor{color1}{rgb}{1,0.498039215686275,0.0549019607843137}
\definecolor{color2}{rgb}{0.172549019607843,0.627450980392157,0.172549019607843}

\begin{axis}[
height=\figureheight,
legend cell align={left},
legend style={fill opacity=0.8, draw opacity=1, text opacity=1, draw=white!80!black},
tick align=outside,
tick pos=left,
width=\figurewidth,
x grid style={white!69.0196078431373!black},
xlabel={\(\displaystyle \lambda\)},
xmajorgrids,
xmin=-1.21, xmax=0.015,
xtick style={color=black},
legend image post style={scale=2},
y grid style={white!69.0196078431373!black},
ylabel={\(\displaystyle \chi\)},
ymajorgrids,
ymin=-1.66349944710653, ymax=34.9334883892372,
ytick style={color=black}
]
\addplot [draw=color0, fill=color0, mark=*, mark size=1, only marks]
table{%
x  y
-1.2 15.8123322720432
-1.18 15.4723665219461
-1.15 15.1231192802352
-1.13 14.7644705333333
-1.1 14.39632871472
-1.07 14.0186371607148
-1.05 13.6313832726401
-1.02 13.2346127458561
-1 12.8284246938965
-0.97 12.4130088092352
-0.95 11.9886387787041
-0.93 11.5557036079745
-0.9 11.1147159248133
-0.88 10.6663448899692
-0.85 10.2114275677701
-0.82 9.7509881278668
-0.8 9.28626039675
-0.78 8.8186964982528
-0.75 8.34997067609653
-0.72 7.8819754660608
-0.7 7.41679406332813
-0.68 6.95667355105333
-0.65 6.5039645458668
-0.63 6.06106340213453
-0.6 5.63032640136333
-0.57 5.21399153589813
-0.55 4.81409673472213
-0.53 4.43240416911053
-0.5 4.07034117075
-0.47 3.72896224896213
-0.45 3.40893049392
-0.42 3.1105277280048
-0.4 2.83367480033333
-0.38 2.57797282776333
-0.35 2.34275677921613
-0.33 2.12714369616653
-0.3 1.93009894533333
-0.28 1.7504761354092
-0.25 1.5870745208748
-0.23 1.43867144836813
-0.2 1.30405574639413
-0.17 1.18205241404813
-0.15 1.07154003788813
-0.13 0.9714622344012
-0.1 0.880834270626133
-0.07 0.7987471733388
-0.05 0.7243666456368
-0.03 0.656932349398533
0 0.5957550830412
0.03 0.5402119035648
0.05 0.4897413879372
0.07 0.4438390602672
0.1 0.40205300268
0.13 0.3639787788972
0.15 0.329253975514133
0.17 0.2975557988172
0.2 0.268595523168133
0.23 0.242117003603333
0.25 0.217890155440133
0.28 0.195711525446533
0.3 0.175399240333333
0.33 0.156792094728533
0.35 0.139746516038533
0.38 0.124134770976133
0.4 0.109843560960133
0.42 0.0967728869292
0.45 0.0002395204608
0.47 0.0145510806145333
0.5 0.0544148579861333
0.53 3.32516352e-05
0.55 0.0149002413132
};
\addlegendentry{$L_x=16$}
\addplot [draw=color1, fill=color1, mark=*, mark size=1, only marks]
table{%
x  y
-1.2 24.5407358965983
-1.18 24.014720941649
-1.15 23.4735552870348
-1.13 22.916950553086
-1.1 22.3446407500808
-1.07 21.7564100751514
-1.05 21.1520845101288
-1.02 20.5315441263931
-1 19.8947595552356
-0.97 19.2417826304348
-0.95 18.5727903056662
-0.93 17.8881006984014
-0.9 17.188205772857
-0.88 16.4738252015834
-0.85 15.7459280632905
-0.82 15.0058079578282
-0.8 14.2551274320783
-0.78 13.4959853943382
-0.75 12.7309697587201
-0.72 11.9632006009214
-0.7 11.1963513100522
-0.68 10.4346394390762
-0.65 9.68275377642609
-0.63 8.94572910440217
-0.6 8.2287713970087
-0.57 7.53701493395904
-0.55 6.87526037972904
-0.53 6.24772004921357
-0.5 5.657788393054
-0.47 5.10788673091339
-0.45 4.59939081574339
-0.42 4.13264846649052
-0.4 3.70706969045435
-0.38 3.32127172455652
-0.35 2.97325747779339
-0.33 2.66059859026304
-0.3 2.38060368354817
-0.28 2.13046795180835
-0.25 1.90738673413704
-0.23 1.70864704411835
-0.2 1.53168164688139
-0.17 1.37410842101009
-0.15 1.23374884848009
-0.13 1.10863342060835
-0.1 0.996995628564261
-0.07 0.897262429567391
-0.05 0.808042304014261
-0.03 0.728105245434783
0 0.656370583313043
0.03 0.591889083556522
0.05 0.533828867639043
0.07 0.481461448153391
0.1 0.434148194617044
0.13 0.391331522020522
0.15 0.352521829306522
0.17 0.317291860800348
0.2 0.285266835022261
0.23 0.256118823204261
0.25 0.229559938654
0.28 0.205338664564261
0.3 0.183234584448783
0.33 0.163054100974
0.35 0
0.38 0.127807202062261
0.4 0.112462158052174
0.42 0.0984784769113913
0.45 0.0857561712587826
0.47 0.0742066851005218
0.5 0.122740028252174
0.53 0.0541462579481739
0.55 0.0181000326282609
};
\addlegendentry{$L_x=24$}
\addplot [draw=color2, fill=color2, mark=*, mark size=1, only marks]
table{%
x  y
-1.2 33.2699889421306
-1.18 32.5579128983546
-1.15 31.824815034848
-1.13 31.0702350281538
-1.1 30.2937401178865
-1.07 29.4949405437446
-1.05 28.6735030728672
-1.02 27.8291575827223
-1 26.9617187642081
-0.97 26.0711086359903
-0.95 25.1573849808532
-0.93 24.2207700421203
-0.9 23.2616965433806
-0.88 22.2808539791693
-0.85 21.2792577622426
-0.82 20.258313405401
-0.8 19.2199164284532
-0.78 18.1665519789461
-0.75 17.1014195474209
-0.72 16.0285405047113
-0.7 14.9528792216983
-0.68 13.8804041069823
-0.65 12.8180966771386
-0.63 11.7738325731872
-0.6 10.7561491695796
-0.57 9.77383680254716
-0.55 8.83543280005806
-0.53 7.94862979864265
-0.5 7.1197400291629
-0.47 6.35326670536265
-0.45 5.65169361799458
-0.42 5.01547706984058
-0.4 4.44324926448316
-0.38 3.93215723062606
-0.35 3.47826541471129
-0.33 3.07695716129032
-0.3 2.72329174154845
-0.28 2.41229254505806
-0.25 2.13914752420961
-0.2 1.68878170242581
-0.17 1.50372514454716
-0.15 1.34088896257606
-0.13 1.19737608173652
-0.1 1.07065771860645
-0.07 0.958538552098645
-0.05 0.859117736497806
-0.03 0.770755792619613
0 0.692038349800258
0.5 0.00458242452258065
0.53 0.0388137261786452
0.55 0.0548845855225807
};
\addlegendentry{$L_x=32$}
\end{axis}

\end{tikzpicture}
    \vspace{-0.1cm}
\caption{Susceptibility as a function of $\lambda$ at $\mu_x = 0$. The phase for large negative
lambda has a broken symmetry, indicated by the increase in $\chi$ with $L_x$. In the symmetry
restored phase, $\chi$ does not increase with the system size as expected.}
\label{fig:chi}
\end{figure}

 For the quantum link ladder with $L_y = 2$ and open boundary conditions in the $x$-direction,
the symmetries are different. In particular, the lattice translation symmetry is not
exact any more, while the charge conjugation symmetry is still exact. Thus, we expect
a phase transition from a phase which breaks charge conjugation symmetry into a phase
where the symmetry is restored. The susceptibility is expected to be large in the symmetry
broken phase and vanish in the phase where the symmetry is restored. \Cref{fig:chi} shows
the expected behaviour of the susceptibility as a function of $\lambda$ for three different
lattice sizes.

\begin{figure}[htp!]
    \centering
    \setlength{\figureheight}{0.38\linewidth}
    \setlength{\figurewidth}{0.95\linewidth}
\begin{tikzpicture}

\definecolor{color0}{rgb}{0.12156862745098,0.466666666666667,0.705882352941177}

\begin{groupplot}[group style={group size=1 by 3,  vertical sep=0.5cm}]
\nextgroupplot[
height=\figureheight,
legend cell align={left},
legend style={
  fill opacity=0.8,
  draw opacity=1,
  text opacity=1,
  at={(0.01,0.95)},
  anchor=north west,
  draw=none
},
minor xtick={},
minor ytick={},
scaled x ticks=manual:{}{\pgfmathparse{#1}},
tick align=outside,
tick pos=left,
width=\figurewidth,
x grid style={white!69.0196078431373!black},
xmajorgrids,
xmin=0.185, xmax=1.52,
xtick style={color=black},
legend image post style={scale=3},
xtick={0.28,0.34,0.46,0.6,0.74,0.86,1,1.12,1.26,1.36,1.48},
xticklabels={},
y grid style={white!69.0196078431373!black},
ylabel={\(\displaystyle W_x\)},
ymin=-0.64716075, ymax=13.59037575,
ytick style={color=black},
ytick={-10,0,10,20}
]
\addplot [color0, mark=triangle*, mark size=1, mark options={solid}]
table {%
0.2 0
0.22 0
0.24 0
0.26 0
0.28 0
0.3 1.251427
0.32 1.251427
0.34 1.251427
0.36 2.235171
0.38 2.235171
0.4 2.235171
0.42 2.235171
0.44 2.235171
0.46 2.235171
0.48 3.125955
0.5 3.125955
0.52 3.125955
0.54 3.125955
0.56 3.125955
0.58 3.125955
0.6 3.125955
0.62 4.032303
0.64 4.032303
0.66 4.032303
0.68 4.032303
0.7 4.032303
0.72 4.032303
0.74 4.032303
0.76 5.00217
0.78 5.00217
0.8 5.00217
0.82 5.00217
0.84 5.00217
0.86 5.00217
0.88 6.058487
0.9 6.058487
0.92 6.058487
0.94 6.058487
0.96 6.058487
0.98 6.058487
1 6.058487
1.02 7.214094
1.04 7.214094
1.06 7.214094
1.08 7.214094
1.1 7.214094
1.12 7.214094
1.14 8.476875
1.16 8.476875
1.18 8.476875
1.2 8.476875
1.22 8.476875
1.24 8.476875
1.26 8.476875
1.28 9.851646
1.3 9.851646
1.32 9.851646
1.34 9.851646
1.36 9.851646
1.38 11.34068
1.4 11.34068
1.42 11.34068
1.44 11.34068
1.46 11.34068
1.48 11.34068
1.5 12.943215
};
\addlegendentry{$L_x=16$}

\nextgroupplot[
height=\figureheight,
minor xtick={},
minor ytick={},
scaled x ticks=manual:{}{\pgfmathparse{#1}},
tick align=outside,
tick pos=left,
width=\figurewidth,
x grid style={white!69.0196078431373!black},
xmajorgrids,
xmin=0.185, xmax=1.52,
xtick style={color=black},
legend image post style={scale=3},
xtick={0.28,0.34,0.46,0.6,0.74,0.86,1,1.12,1.26,1.36,1.48},
xticklabels={},
y grid style={white!69.0196078431373!black},
ylabel={\(\displaystyle \partial E_0/\partial \mu_x\)},
ymin=-22.0637550000001, ymax=1.050655,
ytick style={color=black},
ytick={-40,-20,0,20}
]
\addplot [semithick, color0, mark=triangle*, mark size=1, mark options={solid}]
table {%
0.2 0
0.22 0
0.24 0
0.26 0
0.28 -1.98475000000009
0.3 -1.99999999999996
0.32 -1.99999999999996
0.34 -1.98804999999993
0.36 -4.00000000000027
0.38 -3.99999999999992
0.4 -3.99999999999992
0.42 -3.99999999999992
0.44 -4.00000000000027
0.46 -4.69074999999997
0.48 -5.99999999999987
0.5 -5.99999999999987
0.52 -6.00000000000023
0.54 -5.99999999999987
0.56 -6.00000000000023
0.58 -5.99999999999987
0.6 -7.63715000000005
0.62 -7.99999999999983
0.64 -8.00000000000019
0.66 -7.99999999999983
0.68 -8.00000000000019
0.7 -7.99999999999983
0.72 -8.00000000000019
0.74 -9.94119999999974
0.76 -10.0000000000001
0.78 -10.0000000000001
0.8 -9.99999999999979
0.82 -10.0000000000001
0.84 -9.99999999999979
0.86 -10.2583500000002
0.88 -12.0000000000001
0.9 -11.9999999999997
0.92 -12.0000000000001
0.94 -12.0000000000001
0.96 -12.0000000000001
0.98 -11.9999999999997
1 -12.9312000000002
1.02 -13.9999999999997
1.04 -14.0000000000001
1.06 -14.0000000000001
1.08 -14.0000000000001
1.1 -14.0000000000001
1.12 -14.1147499999999
1.14 -16
1.16 -16
1.18 -16
1.2 -16
1.22 -16
1.24 -16
1.26 -17.8586499999998
1.28 -18.0000000000003
1.3 -18
1.32 -18
1.34 -18
1.36 -18.1626999999999
1.38 -19.9999999999999
1.4 -20.0000000000003
1.42 -19.9999999999999
1.44 -19.9999999999999
1.46 -19.9999999999999
1.48 -21.0131000000001
};

\nextgroupplot[
height=\figureheight,
minor xtick={},
minor ytick={},
tick align=outside,
tick pos=left,
width=\figurewidth,
x grid style={white!69.0196078431373!black},
xlabel={\(\displaystyle \mu_x\)},
xmajorgrids,
xmin=0.185, xmax=1.52,
xtick style={color=black},
xtick={0.28,0.34,0.46,0.6,0.74,0.86,1,1.12,1.26,1.36,1.48},
xticklabels={,0.34,0.46,0.6,0.74,0.86,1,1.12,1.26,1.36,1.48},
y grid style={white!69.0196078431373!black},
ylabel={\(\displaystyle M\)},
ymin=-0.0327295375, ymax=0.6873202875,
ytick style={color=black},
ytick={-0.5,0,0.5,1}
]
\addplot [color0, mark=triangle*, mark size=1, mark options={solid}]
table {%
0.2 0.65459075
0.22 0.65459075
0.24 0.65459075
0.26 0.65459075
0.28 0.65459075
0.3 0
0.32 0
0.34 0
0.36 0.1111061875
0.38 0.1111061875
0.4 0.1111061875
0.42 0.1111061875
0.44 0.1111061875
0.46 0.1111061875
0.48 0
0.5 0
0.52 0
0.54 0
0.56 0
0.58 0
0.6 0
0.62 0.0892721875
0.64 0.0892721875
0.66 0.0892721875
0.68 0.0892721875
0.7 0.0892721875
0.72 0.0892721875
0.74 0.0892721875
0.76 0
0.78 0
0.8 0
0.82 0
0.84 0
0.86 0
0.88 0.055579875
0.9 0.055579875
0.92 0.055579875
0.94 0.055579875
0.96 0.055579875
0.98 0.055579875
1 0.055579875
1.02 0
1.04 0
1.06 0
1.08 0
1.1 0
1.12 0
1.14 0.0308466875
1.16 0.0308466875
1.18 0.0308466875
1.2 0.0308466875
1.22 0.0308466875
1.24 0.0308466875
1.26 0.0308466875
1.28 0
1.3 0
1.32 0
1.34 0
1.36 0
1.38 0.014779625
1.4 0.014779625
1.42 0.014779625
1.44 0.014779625
1.46 0.014779625
1.48 0.014779625
1.5 0
};
\end{groupplot}

\end{tikzpicture}
    \vspace{-0.1cm}
\caption{(Top) Winding number sectors for $L_x=16$. (Middle) The derivative of the ground state
energy with respect to the chemical potential. (Bottom) The magnetization, $M$, defined as the
difference of the plaquettes flippable in the clockwise and anti-clockwise fashion respectively.}
\label{fig:winding_sectors}
\end{figure}
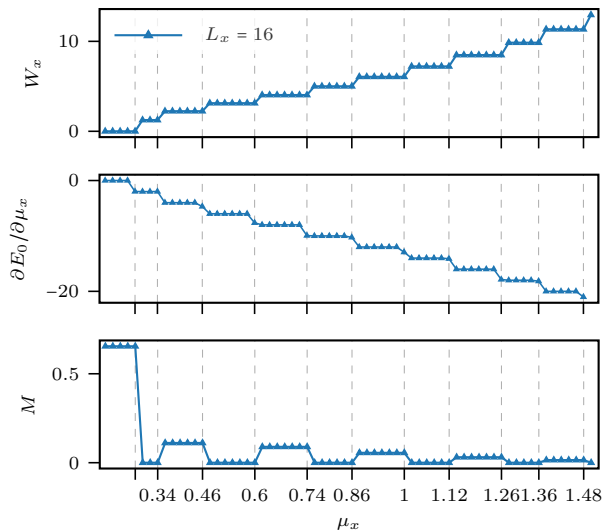

\paragraph{Observables at $\mu_x > 0$.--}
 As the $\mu_x$ is cranked up (keeping $\mu_y=0$), we study the behaviour of the expectation of the winding number 
$\braket{W_x}$, and the derivative of the ground state energy $E_0$ with respect to $\mu_x$,
$\frac{\partial E_0}{\partial \mu_x}$. In the top and the middle panel in \Cref{fig:winding_sectors}, 
we plot the two quantities. As expected from the Feynman-Hellman theorem, these two quantities 
show the same qualitative behaviour upto a constant factor (with an overall negative sign). 
Finally, we also plot the flippability of the plaquettes, $M$. The flippability is defined
as the difference between the total number of plaquettes which are flippable in the clockwise
fashion and the ones which are flippable in the anti-clockwise fashion. Interestingly, this 
observable shows a staggered behaviour between zero and non-zero values for odd and even winding 
number of electric fluxes respectively. This is a clear indication of a co-operative behaviour
of the plaquettes across the lattice.
\end{document}